\documentstyle[11pt,aaspp4]{article}

\def\etal{{\rm et~al. }}
\def\hmpc{\;h^{-1}{\rm Mpc}}
\def\invhmpc{\;h\;{\rm Mpc}^{-1}}

\def\kms{{\rm \;km\;s^{-1}}}
\def\invkms{({\rm km\;s}^{-1})^{-1}}

\def\lya{Ly$\alpha$}

\def\nh1{n_{\rm HI}}

\def\deltaf{\Delta^{2}_{\rm F}(k)} 
\def\obj{\Omega_{b}^{2}/\Gamma}
\def\overdenb{(\rho_b/{\overline\rho}_b)}
\def\overden{(\rho/{\overline\rho})}
\def\K{{\rm K}}
\def\pfk{P_{\rm F}(k)}
\def\p1dk{P_{\rm 1D}(k)}
 
\begin{document}
 
\title{Recovery of the Power Spectrum of Mass Fluctuations from 
Observations of the Lyman-alpha Forest}
\author{Rupert A.C. Croft$^{1,5}$, David H. Weinberg$^{1,6}$,
Neal Katz$^{2,7}$, and Lars Hernquist$^{3,4,8}$}
 
\footnotetext[1]{Department of Astronomy, The Ohio State University,
Columbus, OH 43210}
\footnotetext[2]{ Department of Physics and Astronomy, 
University of Massachusetts, Amherst, MA, 01003}
\footnotetext[3]{Lick Observatory, University of California, Santa Cruz, 
CA 95064}
\footnotetext[4]{Presidential Faculty Fellow}
\footnotetext[5]{racc@astronomy.ohio-state.edu}
\footnotetext[6]{dhw@astronomy.ohio-state.edu}
\footnotetext[7]{nsk@kestrel.phast.umass.edu}
\footnotetext[8]{lars@helios.ucolick.org}
 
\begin{abstract}
We present a method to recover the shape and amplitude of the power spectrum of
mass fluctuations, $P(k)$, from observations of the high redshift \lya\ forest.
The method is motivated by the physical picture that has emerged from 
hydrodynamic cosmological simulations and related semi-analytic models, in 
which typical \lya\ forest lines arise in a diffuse, continuous, fluctuating 
intergalactic medium.  The thermal state of this low density gas 
($\delta\rho/\rho \la 10$) is governed by simple physical processes, which lead
to a tight correlation between the \lya\ optical depth and the 
underlying matter density.  To recover the mass power spectrum, we (1) apply a 
monotonic Gaussian mapping to convert the QSO spectrum to an approximate 
line-of-sight density field with arbitrary normalization, (2) measure the power
spectrum of this continuous density field and convert it to the equivalent 
3-dimensional $P(k)$, and (3) evolve cosmological simulations 
with this $P(k)$ shape and a range of normalizations and choose the
normalization for which the simulations reproduce the observed power
spectrum of the transmitted QSO flux.
Imposing the observed mean \lya\ opacity as a constraint
in step (3) makes the derived $P(k)$ normalization insensitive
 to the choice of 
cosmological parameters, ionizing background spectrum, or reionization 
history.  Thus, in contrast to estimates of $P(k)$ from galaxy clustering, 
there are no uncertain ``bias parameters'' in the recovery of the mass power 
spectrum from the \lya\ forest.
We test the full recovery procedure on smoothed-particle hydrodynamics (SPH) 
simulations of three different cosmological models and show that it 
recovers the true mass power spectrum of the models on comoving scales 
$\sim 1 - 10 h^{-1}$ Mpc, the upper scale being set by the size of the 
simulation boxes. 
The procedure works well even when it is applied to noisy (S/N$\sim$10), 
moderate resolution ($\sim 40\;\kms$ pixels) spectra.  
We present an illustrative application to Songaila \& Cowie's Keck 
HIRES spectrum of Q1422+231; the recovered $P(k)$ is consistent with that of an
$\Omega=1$, $h=0.5$, $\sigma_8(z=0) \approx 0.5$ cold dark matter model.
The statistical uncertainty in this result is large because it is based on a 
single QSO, but the method can be applied to large samples of existing 
QSO spectra and should thereby yield the power spectrum of mass fluctuations
on small and intermediate scales at redshifts $z \sim 2-4$.
\end{abstract}
 
 
\section{Introduction}
\bigskip
A major goal of observational cosmology is the determination of the primordial
power spectrum of mass fluctuations, $P(k)$. This spectrum 
is a direct prediction
of theories of cosmic structure formation, and 
precise measurements of $P(k)$ would allow one to test
these theories and constrain cosmological parameters, especially
when combined with constraints from cosmic microwave
background (CMB) anisotropies on very large scales.
Most efforts to measure $P(k)$ have focused on galaxy redshift surveys. 
This route to the primordial power spectrum has several obstacles, 
including shot noise stemming from the discrete
nature of the galaxy distribution and non-linear gravitational evolution
of $P(k)$, which is important over much of the accessible range of scales. 
The most difficult problem to overcome is the uncertain relation 
between the galaxy and mass distributions, usually parametrized in terms 
of a (possibly scale-dependent) ``bias factor'' between the galaxy and 
mass power spectra.  Furthermore, galaxy redshift surveys primarily
probe structure at a single epoch, redshift zero.
There are studies of clustering evolution, but the noise problems 
that afflict power spectrum measurements are even more severe in dilute,
high redshift samples, and the evolution of bias is uncertain.

Recent theoretical models of the \lya\ forest predict a tight,
physically simple relation between the optical depth for \lya\ absorption 
and the underlying mass density (\cite{bi93}; \cite{bi95};
\cite{reisenegger95}; \cite{bi97}; \cite{cwkh97}; \cite{hui97}).
This paper describes a method that exploits this tight relation to
recover the mass power spectrum in the high redshift universe from 
QSO absorption spectra.
We test the method using cosmological simulations and present an
illustrative application to a single QSO spectrum (of Q1422+231).

Studies of the autocorrelation function of \lya\ forest lines have
shown little evidence for clustering on velocity scales 
$\Delta v \ga 300\;\kms$ and only weak clustering on smaller scales
(\cite{cristiani97} and references therein).
Pando \& Fang (1995), however, find significant large scale clustering in a 
power spectrum analysis of forest lines, using a technique based on wavelets.
Correlation analyses of metal line absorbers also show evidence for
large scale clustering (\cite{sargent89}; \cite{heisler89}; 
\cite{quashnock96}; \cite{quashnock97}).
The approach we take in this paper differs from these earlier studies in 
several respects. First, we work directly
with the observed flux instead of with lines identified from it
(as in the autocorrelation analyses of Press \& Rybicki [1993] and
Zuo \& Bond [1994], though we focus on the power spectrum rather than
the autocorrelation function).  This 
approach has the advantage of reducing shot noise, as information in the 
continuous QSO spectrum is not condensed into a relatively small set
of discrete numbers (line redshifts). It also circumvents ambiguities 
associated with line-identification algorithms, and it can be applied to 
spectra that have insufficient spectral resolution and/or signal-to-noise 
ratio for secure line identification. 
Equally important, we show how to recover the {\it amplitude}
of the mass power spectrum from our measurements. 
The power spectrum of lines would be related to this quantity
by an uncertain, and probably model dependent, ``bias factor''
of forest lines.
 
Our method is motivated by and relies upon the physical picture of the
\lya\ forest that has emerged from hydrodynamic simulations 
of cosmic structure formation (\cite{cen94}; \cite{zhang95}; \cite{hkwm96};
\cite{wb96})
and related analytic models of the intergalactic medium
(\cite{bi93}; \cite{bi95}; \cite{reisenegger95}; Hui \etal 1997; \cite{bi97}).
In this picture, the \lya\ forest arises in highly photoionized gas with
density typically between 0.1 and 10 times the cosmic mean. 
For most of the gas in this density regime, the \lya\ optical depth
obeys the approximate relation 
$\tau \propto \nh1 \propto \rho_{b}^{2} T^{-0.7}$,
where $\nh1$ is the density of neutral hydrogen, $T$ is the temperature
of the gas, and $\rho_{b}$ is the local baryon density. Baryons trace the dark
matter in this density regime, and the interplay between cooling by the 
expansion of the universe and heating by photoionization leads to a 
simple, tight relation between gas density and temperature
(\cite{cwkh97}; \cite{hg97}).
Thus, to a good approximation, the optical depth satisfies
$\tau \propto \rho^{\beta}$, with $\beta \sim 1.5-2$. 
This discussion ignores the effects of peculiar 
velocities, thermal broadening, shock heating, and collisional ionization,
but simulations with all of these effects included retain the tight
relation between $\tau$ and $\rho$ (see figure~2 of
\cite{cwkh97}).

The transmitted flux in a QSO spectrum, $F=e^{-\tau}$, 
is monotonically related to $\rho$ in this approximation.
Fluctuations in the density field along the line of sight to the QSO
produce fluctuations in the absorption, which are seen as the \lya\ forest.
Because the relation between $\tau$ and $\rho$ is fairly simple,
one can extract information about the underlying mass density field 
from the observed flux distribution.  One approach would
be to invert the above relations, deriving $\tau$ from $F$ and
$\rho$ from $\tau$. 
However, the mapping between $\rho$ and $F$ is non-linear
(an exponential of a power law), and it depends on the unknown
constant of proportionality in the $\tau-\rho$ relation. Also,
it is difficult to measure $\tau$ accurately
in saturated regions, where $F \simeq 0$.

Rather than attempt a direct inversion, in this paper we use the fact that the 
primordial density field is expected to have a Gaussian probability
distribution function (PDF), at least in inflationary models
for the origin of fluctuations.
We monotonically map the flux in a QSO spectrum back to a Gaussian density
field, as in Weinberg's (1992) method for recovering primordial 
fluctuations from the observed galaxy distribution.
We measure the 1-dimensional power spectrum $\p1dk$
of this Gaussian density field and convert it to the equivalent
3-dimensional $P(k)$.  At this point we have the shape of the initial 
mass $P(k)$, but since the variance of the Gaussian PDF is not yet known,
we have no information about its amplitude. 
To normalize $P(k)$ we evolve an N-body simulation
using Gaussian fluctuations with the derived $P(k)$ for the initial conditions.
We then use the temperature--density
relation to generate QSO spectra from this simulation,
choosing the photoionization rate to reproduce the observed mean \lya\ opacity.
The amplitude of the {\it flux} power spectrum depends almost exclusively on 
the amplitude of the mass $P(k)$. Therefore, when the amplitude of the flux
power spectrum 
in the simulated spectra matches that measured from the  observations,
the underlying mass $P(k)$ has the correct amplitude. We then have the
normalized $P(k)$ of linear mass fluctuations at the redshift probed by the QSO 
spectrum.

Obviously, there are many assumptions and approximations involved in 
the procedure outlined above,
and it is not obvious a priori that it will work. Most of this
paper will be devoted to testing these assumptions and the method as a
whole. In \S 2 we  explain the method for recovering $P(k)$
in more detail, testing  the steps of  the procedure individually. 
In \S 3 we test the procedure as a whole by attempting to recover 
the mass power spectrum from
simulated observational spectra extracted from  hydrodynamic simulations
of different  cosmological models. In \S 4 we apply the method to
the spectrum of QSO Q1422+231 (provided by A. Songaila and L. Cowie).
In \S 5 we summarize our results and outline directions for
future investigation.

\section{Method}
\subsection{Numerical simulations and physical motivation}
QSO spectra extracted from hydrodynamic cosmological simulations will
play the role of simulated observations, against which we test our procedure.
With these simulated spectra, we can control the physical and ``instrumental''
effects incorporated and test their influence in isolation,
and we know the true mass power spectrum that the method should recover.
As already mentioned, our method for recovering $P(k)$ presumes the
basic picture of the \lya\ forest that emerges from these simulations,
and our tests in this paper depend on its validity.
We will briefly mention tests of the scenario itself in \S 5.

Our simulations use the N-body plus smoothed-particle hydrodynamics
(SPH) code TreeSPH (\cite{hernquist89}; \cite{kwh96}).
We consider three different cold dark matter (CDM) cosmological models;
the simulations are the same ones analyzed by Croft \etal (1997a),
and we refer the reader to that paper for details beyond those
given here.  The first model is ``standard'' CDM (SCDM),
with $\Omega=1$, $h=0.5$ (where $h \equiv H_{0}/100\;\kms\;{\rm Mpc}^{-1}$). 
The power spectrum is normalized so that the rms 
amplitude of mass fluctuations in $8\hmpc$ spheres, linearly
extrapolated to $z=0$, is $\sigma_{8}=0.7$.
This normalization is consistent with
that advocated by White, Efstathiou, \& Frenk (1993)
to match the observed masses of rich galaxy clusters,
but it is inconsistent with the normalization implied by the
COBE-DMR experiment.
Our second model is identical to the first
except that $\sigma_{8}=1.2$. This 
higher amplitude is consistent with the 4-year
COBE data (\cite{bennett96}), and we therefore label the model CCDM.
The third model, OCDM, assumes an open universe with 
$\Omega_{0}=0.4$, $h=0.65$.
This model is also COBE-normalized, with $\sigma_{8}=0.75$ 
(\cite{ratra97}).
The baryon density parameter for all of these models 
is $\Omega_{b}=0.0125 h^{-2}$, a value taken from Walker et al.\ (1991).

We simulate one periodic cube of side length $11.111 \hmpc$ for each model,
using $64^{3}$ collisionless dark matter particles and $64^{3}$
gas particles.  Each simulation was evolved to $z=2$. 
We will deal exclusively with the $z=3$ outputs in this paper.
A uniform photoionization field was imposed
and radiative cooling and heating rates calculated assuming optically thin 
gas, as described in Katz \etal (1996).
QSO absorption spectra were extracted from lines of sight through 
the simulation outputs as described in Hernquist \etal (1996) and
Croft \etal (1997a). 

The simulated spectra exhibit a tight relation between the 
\lya\ optical depth, $\tau$, and the underlying baryon density, $\rho_{b}$, 
for $\overdenb \la 10$.  This relation arises because the temperature
of the low density gas is determined by the interplay between
photoionization heating by the UV background and adiabatic cooling by
the expansion of the universe, leading to a simple 
temperature--density relation that is well approximated by a power law,
\begin{equation}
T = T_{0} \overdenb^\alpha.
\label{eqn:td}
\end{equation}
The constants $T_0$ and $\alpha$ depend on the spectral shape of the UV
background and on the history of reionization; they are likely to lie
in the ranges $4000\;\K \la T_0 \la 15,000\;\K$ 
and $0.3 \la \alpha \la 0.6$ (see \cite{hg97}).
The \lya\ optical depth is proportional to the neutral hydrogen
density $\nh1$ (\cite{gunn65}), which for gas in photoionization
equilibrium is proportional to the density multiplied by the
recombination rate.  For temperatures $T \sim 10^4\;$K, the
combination of these effects yields
\begin{eqnarray}
&\tau \propto \rho_b^2 T^{-0.7} = A\overdenb^\beta, &
\label{eqn:tau} \\
&A = 0.946 \left(\frac{1+z}{4}\right)^6 
\left(\frac{\Omega_b h^2}{0.0125}\right)^2
\left(\frac{T_0}{10^4\;{\rm K}}\right)^{-0.7}
\left(\frac{\Gamma}{10^{-12}\;{\rm s}^{-1}}\right)^{-1}
\left(\frac{H(z)}{100\;\kms\;{\rm Mpc}^{-1}}\right)^{-1}, &
\nonumber
\end{eqnarray}
with $\beta \equiv 2 - 0.7\alpha$ in the range $1.6-1.8$.
Here $\Gamma$ is the HI photoionization rate, and $H(z)$ is the
Hubble constant at redshift $z$.
In this paper we will treat the (observationally uncertain) quantity
$\Gamma$ as a free parameter, which we set 
by requiring that our simulated spectra match the observed
mean \lya\ flux decrement at $z=3$, 
$D_A \equiv 1-\langle F \rangle = 0.36$
(\cite{press93}; \cite{rauch97}; but see \cite{ZL93}, who 
estimate a lower mean decrement).
If we adopted a different baryon density or reionization history
in the simulations then the required value of $\Gamma$ would be
different.  As equation~(\ref{eqn:tau}) suggests, in a given 
cosmological model the mean
flux decrement basically constrains the parameter combination
$\Omega_b^2 \Gamma^{-1} T_0^{-0.7}$, and for a fixed
value of this combination one gets nearly identical spectra
from an underlying density field, whatever the individual
values of the parameters. 

The optical depth depends on the density of neutral atoms in
redshift space rather than in real space, so peculiar velocities
and (to a lesser extent) thermal broadening introduce scatter in
the relation between $\tau$ and $\rho_b$.
Equation~(\ref{eqn:tau}) breaks down more drastically in regions with
$\overdenb \ga 10$ because of shock heating and collisional ionization,
but these regions have a small volume filling factor so they affect
only a small fraction of a typical spectrum.
As shown in figure~2 of Croft \etal (1997a), the correlation
between $\tau$ and $\rho_b$ remains tight over most of the spectrum
even when all of the relevant physical effects are taken into account.
Finally, because the gaseous structures responsible for typical \lya\
forest lines are large ($\ga 100$ kpc), low density, and fairly cool
($T \sim 10^4\;$K), pressure gradients have only a 
small effect on their
dynamics relative to gravity, 
and the gas within them traces the underlying dark matter
except on very small scales.  Equation~(\ref{eqn:tau}) therefore provides
a link between the \lya\ optical depth and the total mass density
$\rho = (\Omega/\Omega_b)\rho_b$.  We will show below that, as a result of
this link, the power spectrum of the flux has the same shape as the power
spectrum of the underlying mass fluctuations on large scales.

\subsection{Recovery of the power spectrum shape}

Given the relation between optical depth and density, we would like to 
recover the density field along the line of sight and measure its power 
spectrum. We could attempt to recover $\rho$ by inverting 
equation~(\ref{eqn:tau}), but we would encounter several obstacles.
In high density regions the QSO spectrum saturates, and when $F \approx 0$
it is difficult to measure $\tau = -{\rm ln}\; F$ accurately.
Since the power spectrum weights high amplitude regions strongly
(the galaxy power spectrum, for instance, is strongly affected by galaxies
in rich clusters even though field galaxies are much more common), 
small amounts of noise in saturated regions could produce large uncertainties 
in the power spectrum of the recovered density field.
Obtaining $\rho$ by direct inversion also requires
a priori knowledge of the
proportionality constant $A$ in equation~(\ref{eqn:tau}).
A given spectrum could potentially be produced
either by a weakly fluctuating density field with a large value of $A$
or by a strongly fluctuating density field with a small value of $A$.
Since the $\tau-\rho$ relation is non-linear, changes in $A$
are not equivalent to linear rescalings of $P(k)$.
When we extract spectra from simulations, we fix $A$ by matching
the observed mean flux decrement, but we are able to do so only because
the simulation itself provides the density field. Finally, from the
point of view of testing cosmological models, we are interested
primarily in the power spectrum of the initial, linear mass 
fluctuations, rather than the power spectrum of the non-linear density
field.  At large scales and high redshifts, non-linear evolution has
little effect on $P(k)$, and the linear and non-linear power spectra
can be related by analytic approximations (\cite{hamilton91};
\cite{peacock96}; \cite{jain95}) or by numerical calculations.
However, to the extent that we can recover the linear $P(k)$ directly from
the observations, we are somewhat closer to our ultimate goal of
addressing cosmological questions.

We can circumvent all of these obstacles --- saturation, the
unknown value of $A$, and non-linear evolution of the density field ---
using ``Gaussianization,'' a technique introduced by Weinberg (1992)
as a tool for recovering primordial density fluctuations from
the observed galaxy distribution.  Inflationary models for the origin
of structure predict that the initial density field has a Gaussian PDF.
Gravitational evolution skews the PDF, but it tends to carry high
density regions of the initial conditions into high density regions
of the evolved density field, low density regions into low density regions,
and so on in between; it therefore preserves an approximately monotonic
relation between initial and final density even on scales that are 
moderately non-linear.  Given an evolved density field on a grid,
one can recover an approximate map of the initial fluctuations by
sorting the pixels in order of density and then assigning new densities
to the pixels so that they have the same rank order but a Gaussian PDF.
The overall amplitude of the fluctuations, corresponding to the
width of the Gaussian, must be determined in a separate step,
by evolving the recovered initial conditions and comparing them to
the observations.

In the application described by Weinberg (1992), the input data
set is a galaxy density field from a redshift survey.
In our case we will apply Gaussianization to \lya\ forest spectra,
where the physical discussion in \S 2.1 gives us excellent reason
to expect a monotonic relation between observed flux and mass
density, approximately $F = {\rm exp}[-A\overden^\beta].$
Because Gaussianization uses only the rank order of the pixel densities,
we do not need to know the parameters of this non-linear relation,
only that it is monotonic.
An individual spectrum yields only a 1-dimensional probe through
the density field, but we aim to recover
the power spectrum of fluctuations rather than the full density field.
Peculiar velocities can depress the power spectrum on small scales
by smoothing structure in redshift space and adding scatter to the
$\tau-\rho$ relation. However, on large scales they should not alter the
shape of $P(k)$.  Coherent flows into high density regions and out
of low density regions can amplify the redshift-space
power spectrum (\cite{kaiser87}), but our normalization procedure
(described in \S 2.3 below) will automatically account for this effect.

\begin{figure}
\centering
\vspace{12.5cm}
\includegraphics{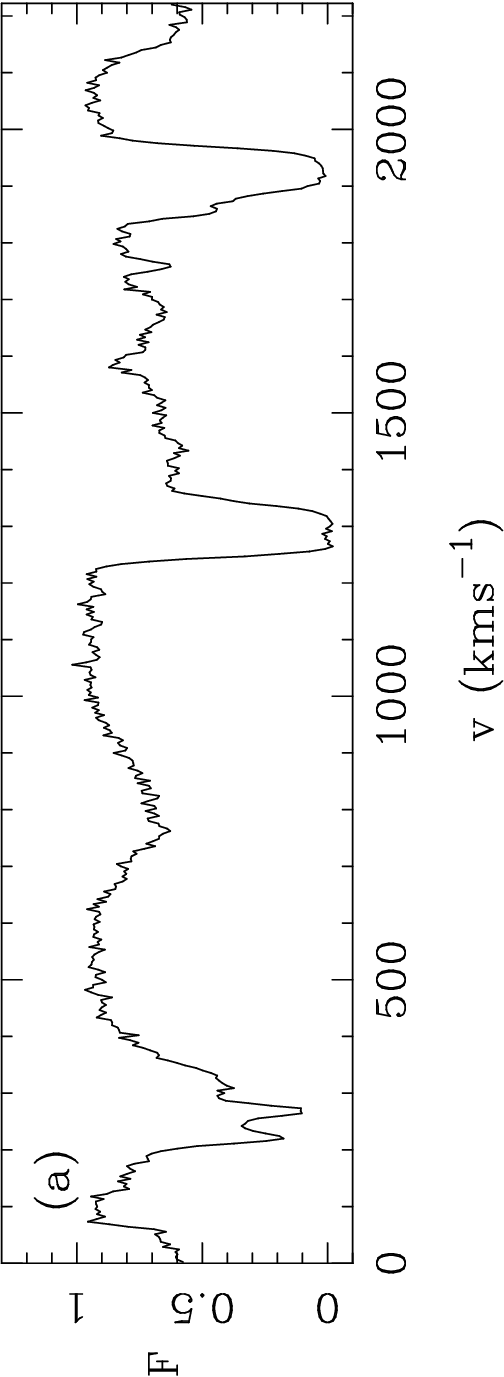}
\includegraphics{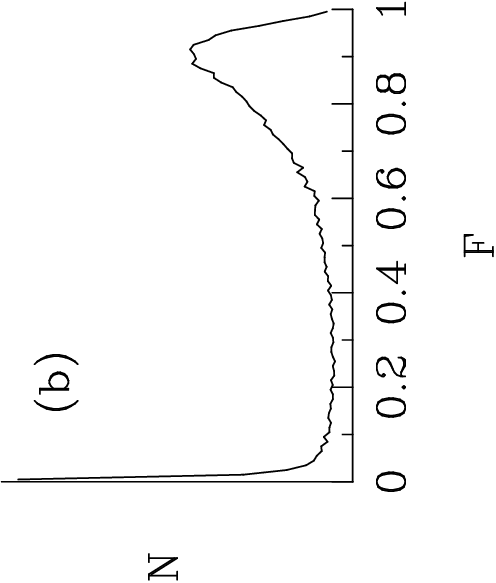}
\includegraphics{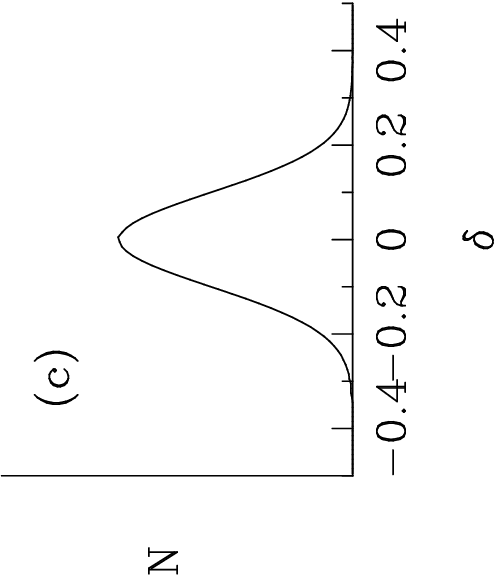}
\includegraphics{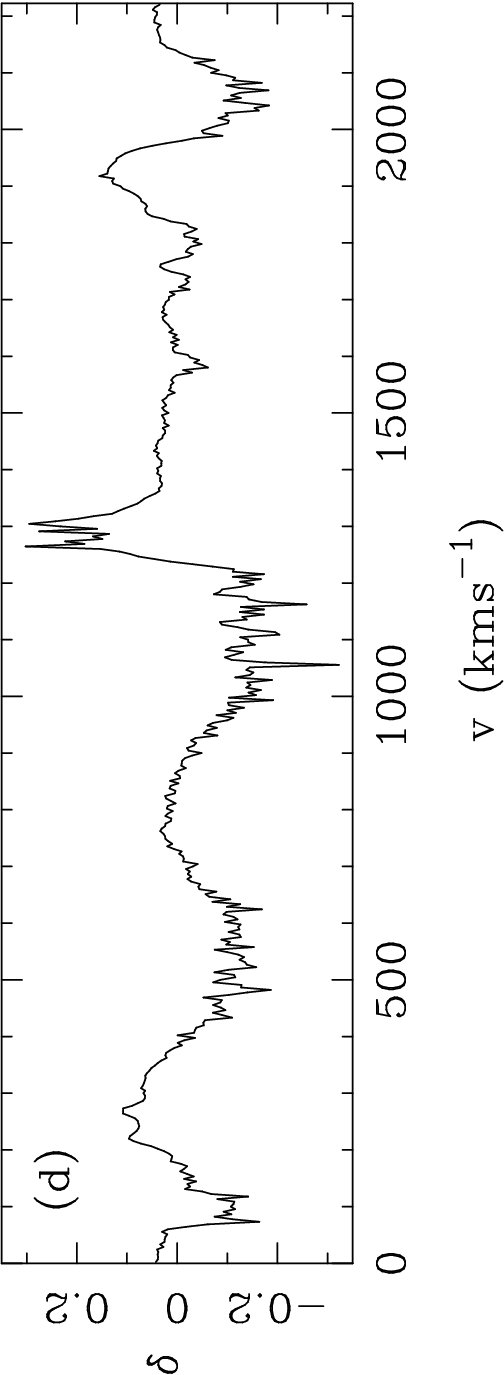}
\caption[junk]{\label{gausshow}
Recovery of a line-of-sight initial density field
from a QSO \lya\ spectrum by Gaussianization.
(a) An absorption spectrum taken from 
a TreeSPH simulation of the SCDM model.
Simulated photon noise (S/N=50 in the continuum) has been added. 
(b) The PDF of pixel flux values in a set of 100 simulated spectra from the
same model. 
(c) The PDF of the inferred initial density contrasts $\delta$.
The pixels in the simulated spectra are rank ordered according to their
flux values, and each pixel is assigned a density contrast such that the 
original rank order is retained and $P(\delta)$ is Gaussian.
The width of the Gaussian, proportional to the
amplitude of the density fluctuations, is arbitrary at this
point and will be determined later by the normalization process
described in \S 2.3. 
(d) The inferred initial density contrast field  along the same 
line of sight as the spectrum in (a).
}
\end{figure}

Figure \ref{gausshow} illustrates the Gaussianization procedure for
a sample spectrum taken from our SCDM model. 
The transmitted flux $F$ (normalized to $F=1$ for no absorption)
is computed along a random line of sight through the simulation.
The observed wavelength is related to the velocity by
$\lambda = \lambda_\alpha (1+z)(1+v/c)$, where $\lambda_\alpha=1216\AA$
is the \lya\ rest wavelength and $z=3$.  The velocity range
$\Delta v = 2222\;\kms$ ($\Delta z = 0.0296$)
is set by the size of the periodic simulation box;
a simulated spectrum like that in Figure~\ref{gausshow}a corresponds
to a small section of an observed QSO spectrum.
We have added noise to
the spectrum (S/N=50 in the unabsorbed regions) in a manner described in 
\S 2.4 below. 
Figure~\ref{gausshow}b shows the PDF of the flux for 100 simulated spectra.
This PDF is mapped by Gaussianization to the Gaussian PDF of the 
density contrast, $P(\delta)$, shown in Figure~\ref{gausshow}c.
Figure~\ref{gausshow}d shows the line-of-sight density contrast
field $\delta(v)$ inferred from the spectrum.
This field is equivalent to the  ``initial'' density contrast, in the sense
that the fluctuations are Gaussian instead of having a PDF skewed by
gravitational evolution. The amplitude of the field is arbitrary at 
this stage, and will be determined by a separate normalization process
(described in \S 2.3).  
Noise in the spectrum causes the small scale spikiness seen in the recovered
density contrast field, especially in the lowest and highest density regions.
Smoothing the spectrum over a few pixels prior to Gaussianization would
suppress the noise and remove this spikiness, but we have not incorporated such
smoothing in our analysis because we find that the noise does not change
the derived power spectrum on the scales where our recovery method is reliable.

Now that we have an approximation to the
initial mass density field along the line of sight, we 
would like to measure the 3-dimensional power spectrum, $P(k)$. 
We first estimate the 1-dimensional power spectrum, 
$\p1dk=\langle \delta^{2}(k) \rangle$, where 
\begin{equation}
\delta(k)=\frac{1}{2\pi}\int\delta(x)e^{-ikx} dx.
\end{equation}
For the simulated spectra, we compute $\delta^{2}(k)$ using a Fast Fourier 
Transform, then average over multiple lines of sight to find $\p1dk$.
When we analyze the spectrum of Q1422+231 in \S 4, we will average over
bins in $k$ rather than multiple lines of sight.
The 1-dimensional power spectrum $\p1dk$ is related to the 
3-dimensional power spectrum $P(k)$ by 
\begin{equation}
\p1dk=\frac{1}{2\pi}\int_{k}^{\infty}P(y) y dy
\label{eqn:p1d}
\end{equation}
(\cite{kaiser91}).
Equation~(\ref{eqn:p1d}) shows that the 1-dimensional power spectrum
on large scales receives contributions from 3-dimensional fluctuations
at all smaller scales (higher $k$).
Because of this aliasing of power, the 1-dimensional power spectra
of pencil beam galaxy redshift surveys can sometimes show spurious
large scale features that are not present in the 3-dimensional power spectrum
(\cite{kaiser91}; \cite{baugh96}).
In order to recover $P(k)$ from $\p1dk$, one must invert 
equation~(\ref{eqn:p1d}),
\begin{equation}
P(k)=-\frac{2\pi}{k} \frac{d}{dk}\p1dk.
\label{eqn:invert}
\end{equation}
The dilute sampling in pencil beam galaxy surveys can make such an 
inversion noisy, but in our case we start from a continuous spectrum
instead of a discrete distribution of objects, and the 
inversion~(\ref{eqn:invert}) proves quite practical even given realistic
levels of observational noise.

\begin{figure}[t]
\centering
\vspace{10.0cm}
\includegraphics{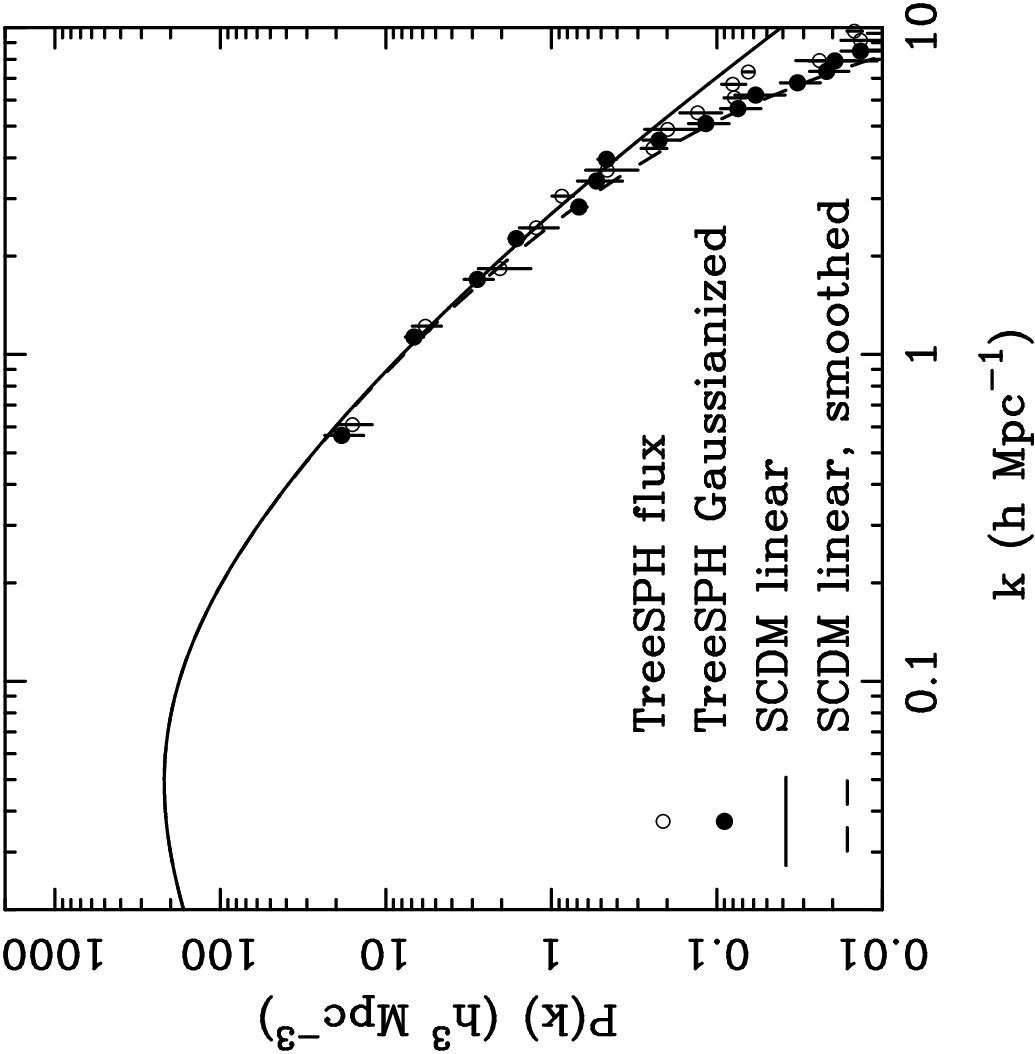}
\caption[junk]{\label{pkscdm}
Recovery of the power spectrum shape from simulated spectra extracted
from the SCDM simulation at $z=3$, with $k$ in comoving $\invhmpc$.
Filled circles show $P(k)$ derived from the Gaussianized flux,
while open circles show the flux power spectrum itself.
Error bars represent the $1\sigma$ dispersion among five sets
of 20 lines of sight through the simulation, each set
roughly equal in redshift path length to an observed QSO spectrum.
The derived power spectra are normalized on large scales
to match the linear theory SCDM mass power spectrum, shown by the solid line.
The dashed line shows the linear theory power spectrum multiplied
by $\exp(-\frac{1}{2}k^2 r_s^2)$, with $r_s=1.5\hmpc$.
}
\end{figure}

Figure~\ref{pkscdm} illustrates the recovery of the shape of $P(k)$
from 100 spectra drawn from the SCDM simulation at $z=3$.
We did not add noise to these spectra; we will examine the effects
of noise and instrumental resolution in \S 2.4 below.
Filled circles represent the 3-dimensional $P(k)$ recovered from
the Gaussianized spectra.  The normalization has been chosen to match 
the amplitude of the linear theory SCDM mass power spectrum at $z=3$,
shown by the solid curve.
Open circles show the 3-dimensional $P(k)$ recovered directly from the
flux, without Gaussianization, again normalized to match
the linear theory mass power spectrum at large scales.

The power spectrum derived from the Gaussianized flux matches the shape of
the linear theory $P(k)$ quite well for comoving wavelengths
$\lambda=2\pi/k \ga 1.5 \hmpc$, up to the scale of the simulation box,
$\lambda=11.111 \hmpc$.  The redshift path length of the \lya\ forest
region for a $z=3$ QSO corresponds to roughly 20 of our simulated
spectra.  We have therefore divided our 100 Gaussianized spectra into
five sets of 20 and computed $P(k)$ separately for each set.
The error bars on the filled circles in Figure~\ref{pkscdm} show the
$1\sigma$ dispersion of these five $P(k)$ estimates and thus indicate
the uncertainty in $P(k)$ expected from a single observed QSO 
spectrum.  Note, however, that our 100 spectra are all drawn from the
same simulation volume and may therefore underestimate the ``cosmic variance''
caused by large scale variations in the density field.  Also, we have
not yet included the effects of instrumental noise and continuum fitting,
nor have we indicated the uncertainty in the overall amplitude of 
$P(k)$ that will arise from our normalization procedure.

On small scales, the recovered $P(k)$ falls below the linear theory
power spectrum. 
The depression of small scale power is caused by the ``smearing''
of the absorption spectrum by peculiar velocities and thermal broadening
and by non-linear gravitational effects that are not fully reversed by 
Gaussianization.
To indicate the scales affected, the dashed line
in Figure~\ref{pkscdm} shows the linear theory SCDM spectrum multiplied
by $\exp(-\frac{1}{2}k^2 r_s^2)$ with $r_s=1.5\hmpc$.  
This smoothed linear theory
spectrum is a good match to the $P(k)$ derived from the Gaussianized flux.
It might be possible to extend the dynamic range of our method by
incorporating analytic (or numerical) corrections for this loss of small
scale power; however, these corrections might well depend on the
assumed cosmological model, and we will not attempt them here.

The power spectrum of the flux also has the same shape as the mass power
spectrum on large scales, even though the relation between flux and
mass density is highly non-linear.  This agreement is reminiscent of
Weinberg's (1995) results for locally biased galaxy formation, which
show that a non-linear but local relation between galaxy and mass density
does not change the shape of the power spectrum on large scales,
though it may change the amplitude by a constant factor.
The error bars, again derived from the dispersion among five sets
of 20 spectra, are slightly larger than those for the Gaussianized flux $P(k)$,
probably because Gaussianization ``regularizes'' the spectra and 
reduces the impact of rare, strong absorption regions.
However, the agreement between the filled and open circles indicates
that the recovery of the power spectrum shape is not
sensitive to our assumption of a Gaussian primordial PDF --- we could
recover the shape of $P(k)$ without Gaussianizing at all, at the cost
of slightly larger statistical uncertainty.
The Gaussian assumption will play a more important role when we 
normalize $P(k)$ as described in \S 2.3 below, using simulations
with random phase initial conditions.

\begin{figure}[t]
\centering
\vspace{10.0cm}
\includegraphics{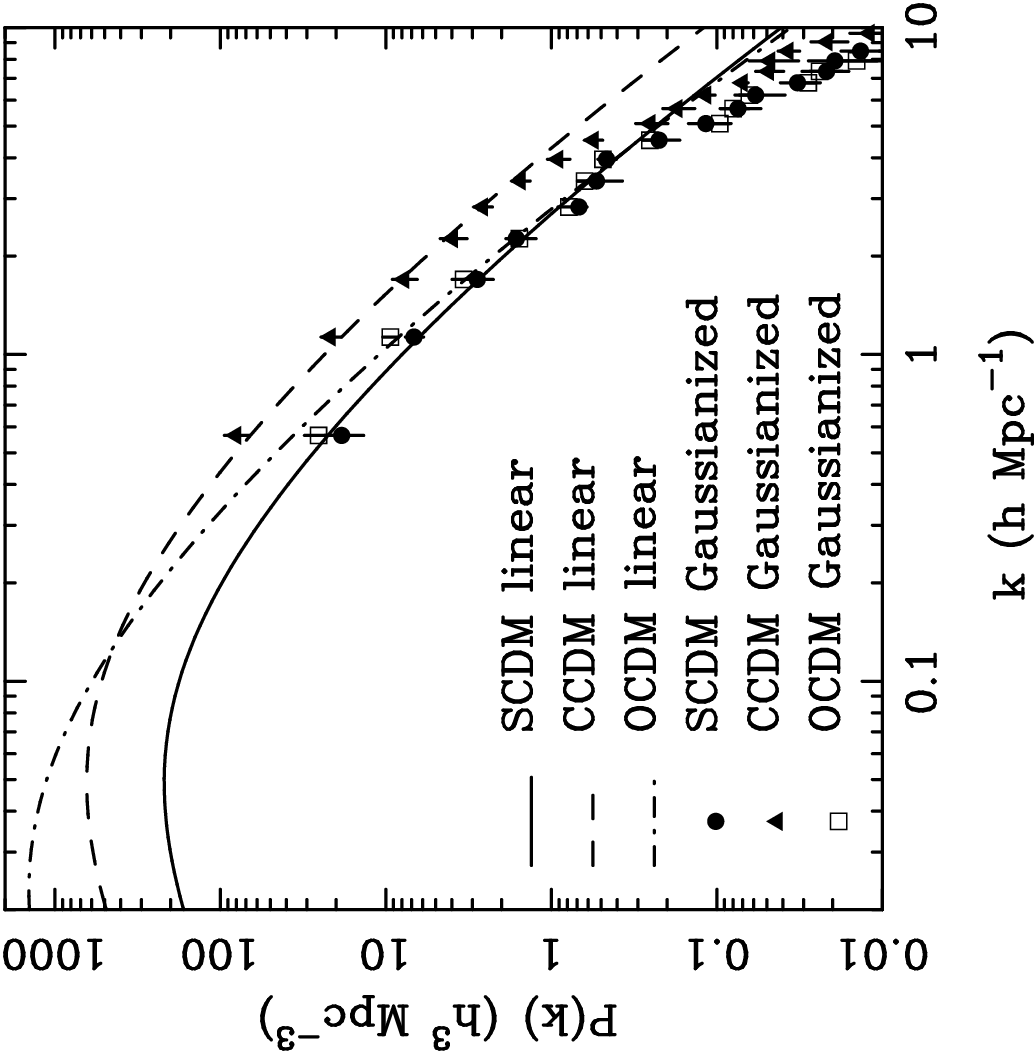}
\caption[junk]{\label{pkgauss}
Recovery of the power spectrum shape for three different CDM models.
Points show $P(k)$ derived from Gaussianized simulated spectra, 
normalized on large scales to match the corresponding linear theory
power spectra shown by the solid (SCDM), dashed (CCDM), and dot-dashed
(OCDM) curves.  See Fig.~\ref{pkscdm} caption for further details.
In all three cases, the shape of the initial power spectrum is recovered 
quite accurately for $0.5 \leq k \leq 4\;\invhmpc$.
}
\end{figure}

Figure~\ref{pkgauss} shows the recovery of the power spectrum shape
for the three different cosmological models (SCDM, CCDM, OCDM)
for which we have TreeSPH simulations.  In each case the normalization
is adjusted to match the linear theory $P(k)$ at large scales.
Results for the SCDM model are repeated from Figure~\ref{pkscdm}.
The power spectrum shape is also correctly recovered for the higher
amplitude, CCDM model, though the drop below linear theory shifts to
somewhat lower $k$ because of the larger scale of non-linearity.
We even recover the subtle difference in shape between the OCDM
and SCDM power spectra, though this difference is even smaller when
the wavenumber is expressed in directly observable units of $(\kms)^{-1}$
(see \S 4), so a clean distinction will require measurements to larger scales.

\begin{figure}[t]
\centering
\vspace{10.0cm}
\includegraphics{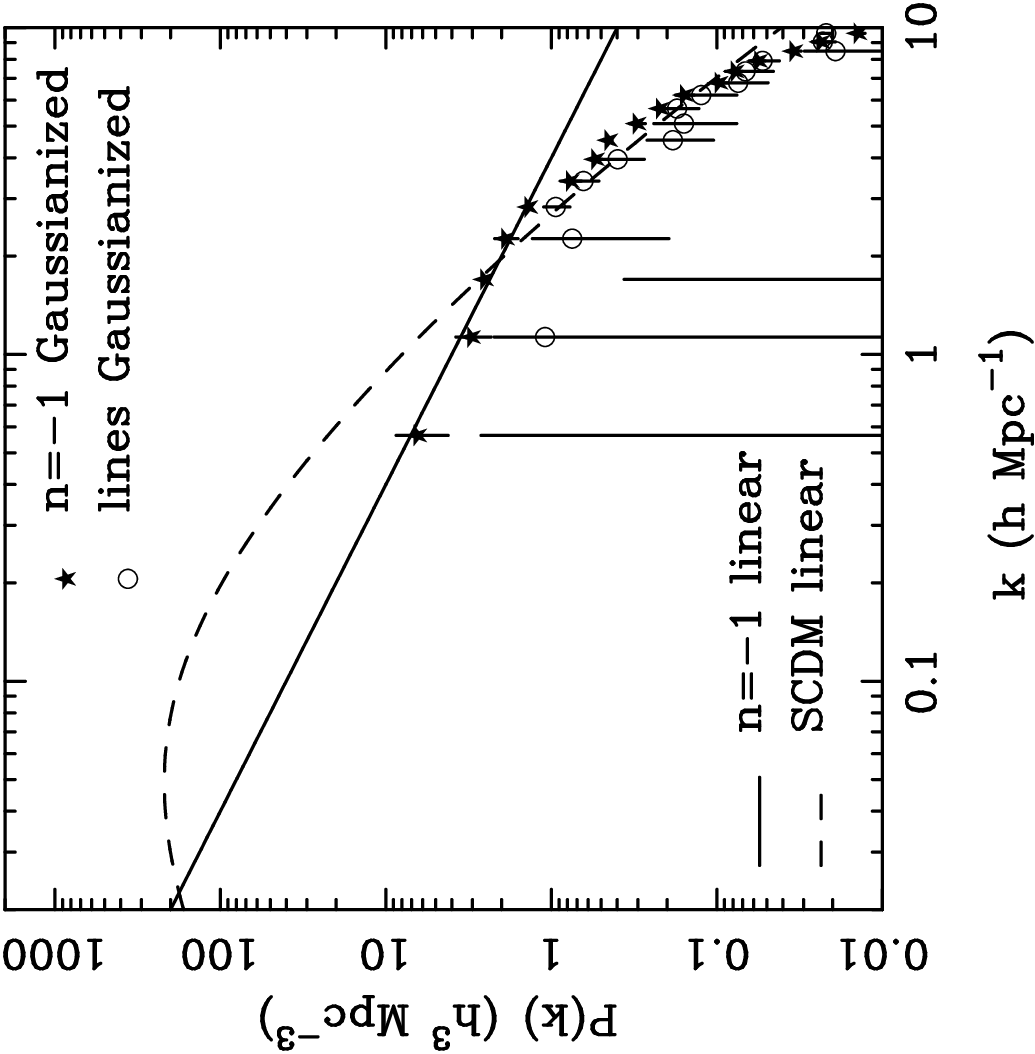}
\caption[junk]{\label{pklinen-1}
The power spectrum of the Gaussianized flux for two non-CDM models.
Stars show results for a model with an $n=-1$ power law initial spectrum
(the linear theory $P(k)$ is shown as a solid line).
Open circles show results from a model where spectra are built
from a random superposition of discrete, unclustered lines with 
Voigt profiles.  The linear theory SCDM power spectrum is shown for comparison.
}
\end{figure}

Since the three CDM models have similar power spectrum shapes, we want
to check that our method can indeed recover the correct shape for
a model with a significantly different $P(k)$.  Figure~\ref{pklinen-1}
shows the $P(k)$ recovered from the Gaussianized flux in a model with
an $n=-1$ power law initial power spectrum (and $\Omega=1$, $h=0.5$).
We do not have a TreeSPH simulation of this model; instead we created
artificial spectra by an extended N-body simulation technique described 
in \S 2.3 below.  At small scales the recovered $P(k)$ is depressed
by non-linear evolution and velocity smoothing and is similar to that
of the CDM models.  However, at $k \la 2.5 \invhmpc$ the recovered
$P(k)$ bends sharply and matches onto the correct linear theory power
spectrum shape, shown by the solid line.  Thus, the method is
clearly capable of distinguishing an $n=-1$ power law model from a 
CDM model.  

As a further test, we have measured the Gaussianized flux $P(k)$ for a model
designed to have no large scale clustering. We use
a program written by J. Miralda-Escud\'{e} that builds
up simulated QSO spectra  by placing discrete absorption
lines with Voigt profiles at random redshifts. 
The HI column densities of the lines are drawn from a power law distribution
$f(\nh1) \propto \nh1^{-1.5}$ and the 
$b$ parameters from a truncated Gaussian distribution peaking at $b=28\;\kms$.
The model and parameters are described in more detail in Croft \etal (1997a). 
The derived power spectrum is shown by the open circles in 
Figure~\ref{pklinen-1}.  On small scales we
see a ``clustering'' signal in the Gaussianized flux caused by the 
finite width of the lines (\cite{zuo94}).
However, the derived $P(k)$ is consistent with zero for all 
$k < 2 \invhmpc$, as expected given the absence of any intrinsic
clustering in the line model.  The Gaussianized flux power spectrum
can clearly distinguish an unclustered line model from a model 
where the \lya\ forest arises in an intergalactic medium with large
scale fluctuations.

\subsection{Power spectrum normalization}
Once we have applied the procedure described in \S 2.2 to simulated
or observed QSO spectra, we know the (approximate) shape of the initial
power spectrum, but we do not know its amplitude.
To determine this normalization, we use the derived $P(k)$ to set
up initial conditions for a series of cosmological simulations,
with  different linear fluctuation amplitudes.
We assign random complex phases to the individual Fourier modes,
again exploiting the theoretical expectation that the primordial
fluctuations are Gaussian.  We evolve these simulations to the observed
redshift and extract artificial spectra.
In simulations with a low fluctuation amplitude, the intergalactic medium
(IGM) is relatively smooth, and the extracted spectra show weaker
fluctuations than the input spectra.  For a high fluctuation amplitude,
the simulated IGM is clumpy, and the corresponding \lya\ spectra
have too much structure.  We take the correct linear theory amplitude
to be the one for which the extracted \lya\ spectra have the correct
degree of structure --- specifically, we normalize the mass $P(k)$
by requiring that the evolved simulations reproduce the 3-dimensional
power spectrum of the (non-Gaussianized) flux on large scales.

This normalization procedure would be computationally impractical
if we had to evolve full hydrodynamic simulations for each trial
fluctuation amplitude.  Fortunately, the physical discussion of the
\lya\ forest in \S 2.1 suggests a useful shortcut.  We use a particle-mesh
(PM) N-body code (\cite{hockney81}) to evolve a collisionless dark
matter simulation from the specified initial conditions.  We assume that 
the baryons trace the dark matter distribution and that the gas
temperature is given by the power law temperature--density 
relation~(\ref{eqn:td}).  We can then extract \lya\ spectra from the
simulated gas distribution in the usual fashion and measure their
flux power spectra.

This simple, ``pseudo-hydro'' technique (details given below) is very
similar to semi-analytic IGM models that have been used to
predict properties of the \lya\ forest (\cite{bi93}; \cite{bi95};
\cite{reisenegger95}; \cite{gnedin96}; \cite{bi97}; \cite{hui97}),
except that these use the lognormal approximation (\cite{coles91})
or variations of the Zel'dovich approximation (\cite{zeldovich70})
to compute the non-linear dark matter density field.
The PM method, being fully non-linear, should yield more accurate
results at a still--modest computational cost.
A similar numerical approach has been used by 
Petitjean \etal (1995; see also \cite{mucket96}) and by
Gnedin \& Hui (1997; see also \cite{gnedin97}); both of these groups also 
incorporate simplified hydrodynamic effects in their PM evolution.
We have visually compared spectra computed by the PM approach to
spectra produced by TreeSPH from the same initial conditions, and we
find good agreement in nearly all regions.
The PM technique breaks down in high density regions where shock heating
and/or radiative cooling drive gas away from the 
temperature--density relation~(\ref{eqn:td}),
which holds only when photoionization heating and adiabatic cooling are
the dominant processes affecting the gas temperature.
In visual comparisons of our present simulation spectra, we see that 
these effects appear to be more important than the effects of
pressure gradients in limiting the accuracy of the approximation.
Our tests below will show that the approximation is
adequate for our present purpose, determining the normalization
of $P(k)$. Clearly this approximation (or the 
``hydro-PM'' approximation of \cite{gh97}) is potentially useful
for other \lya\ forest applications, though its suitability
needs to be tested against full hydrodynamic calculations
on a case-by-case basis.

In addition to the linear fluctuation amplitude, a number of 
uncertain parameters must be specified before a given set of 
initial conditions can be evolved and used to create artificial spectra.
These include the cosmological parameters $\Omega_0$, $\Lambda_0$, and $h$,
and four parameters that influence the density, temperature,
and ionization state of the IGM: $\Omega_b$, $T_0$, $\alpha$, and
$\Gamma$.
One might worry that the derived normalization of $P(k)$ would be
sensitive to the assumed values of these parameters.
Fortunately, the discussion in \S 2.1 suggests that these parameters 
collectively influence the \lya\ spectrum mainly through a
single combination, the constant $A$ in the $\tau-\rho$ 
relation~(\ref{eqn:tau}) (the index $\beta$ has only a slight 
dependence on these parameters).  
For a specified initial $P(k)$, the statistical properties of the underlying
density field depend mainly on the linear fluctuation amplitude at the
redshift in question and are insensitive to cosmological parameters.
Once the PM code has provided the density field, the value of $A$ can
be determined by matching the observed mean flux decrement $D_A$, a
constraint that is independent of the flux power spectrum.
Our tests below will show that once this mean decrement constraint is
imposed the $P(k)$ normalization derived by matching the flux power
spectrum is virtually independent of the uncertainties in
cosmological and IGM parameters.

Our full procedure for 
producing spectra from the PM code is as follows:\\
\noindent (a) We generate initial conditions.
For the tests in this Section, we want to suppress the statistical 
fluctuations caused by variations of structure from one simulation
to another, and we therefore start our PM simulations from the
same initial density fields that were used for the TreeSPH simulations,
varying only the linear fluctuation amplitude.  
In \S 3 and \S 4, where we present end-to-end tests of our method
and apply it to observations, we generate Gaussian fluctuations
with the $P(k)$ shape derived from the input data,
using the  standard technique
of drawing the real and imaginary parts of each Fourier mode from
independent Gaussian distributions with mean zero and variance $P(k)/2$.
In all cases we use the same box size as the TreeSPH simulations 
(11.111 $\hmpc$ comoving) and $64^{3}$ particles.  \\
\noindent (b) We evolve the models from $z=15$ to $z=3$
in 20 timesteps (each timestep corresponding to a change in expansion
factor $\Delta a=0.2$).
We use a $128^3$ mesh for density--potential computations in the
PM code. We must adopt values of $\Omega_{0}, \Lambda_0$, and $h$
for this evolution, but we will show below that the results are
insensitive to this choice.\\
\noindent (c) We interpolate the density and velocity fields onto 
a $128^{3}$ grid using a cloud-in-cell (CIC) scheme. We then smooth the fields
with a Gaussian filter of radius 1 grid cell (following Hui \etal 1997)
in order to ensure that velocities are defined everywhere.
This scheme is different from the method used to extract line-of-sight fields
from the TreeSPH simulations, which involves Lagrangian
smoothing with the SPH kernel (see \cite{hkwm96}). 
The success of our tests below indicates that the flux power
spectrum is insensitive to the detailed procedure used to extract
\lya\ spectra from a simulation.\\
\noindent (d) We select random lines of sight through the simulation
along which to extract spectra.  We assign 
temperatures to each pixel along these lines of sight
using the relation $T=T_{0}\rho^{\alpha}$.
Typical values would be $T_0 \approx 6000\;\K$, $\alpha=0.6$,
but we will show below that the results do not change for other
reasonable choices of $T_0$ and $\alpha$.\\
\noindent (e) We adopt provisional values of $\Omega_b$ and $\Gamma$
and compute the neutral hydrogen fraction in each pixel, using its
density and temperature and assuming photoionization equilibrium.
We compute the \lya\ optical depth $\tau$ from the neutral hydrogen density.\\
\noindent (f) We map the absorption spectrum $\tau(\lambda)$ from real space 
to redshift space, redistributing $\tau$ values as implied by the peculiar
velocity field and convolving with the appropriate thermal broadening.\\
\noindent (g) Finally, we multiply the $\tau$ values in all of the
spectra by a constant chosen so that the mean flux decrement in the
spectra matches the observed $D_A$.  This step is equivalent to
changing the parameter combination $\Omega_b^2/\Gamma$ from the
provisional value adopted in step (e).\\

We compute the flux power spectrum as described in \S 2.2,
estimating $\p1dk$ by FFT and deriving the 3-dimensional
flux power spectrum, $\pfk$, using equation~(\ref{eqn:invert}).
The 3-dimensional $\pfk$ provides a more robust normalization
constraint than the 1-dimensional flux power spectrum because
we can match power on large scales; $\p1dk$ contains aliased small
scale power even at low $k$ (eq.~[\ref{eqn:p1d}]).
In our flux power spectrum plots we show the quantity 
$\deltaf \equiv k^3\pfk$, the contribution of logarithmic $k$ 
intervals to the variance (\cite{peebles80}), instead of $\pfk$.
Because the flux power spectra are steep at the scales that
we examine, the $k^3$ factor reduces the dynamic range of the plots
and makes it easier to discern differences in amplitude.
The use of $\deltaf$ also minimizes potential confusion with
our plots of the recovered mass fluctuation power spectrum,
where we will continue to show $P(k)$ itself.
We note here that the flux power spectrum, $\deltaf$, is an interesting
statistic in its own right, and can be used as a measure of
structure in the \lya\ forest.  However, in this paper we will concern
ourselves solely with its usefulness in determining the normalization
of the mass power spectrum $P(k)$.

\begin{figure}[t]
\centering
\vspace{10.0cm}
\includegraphics{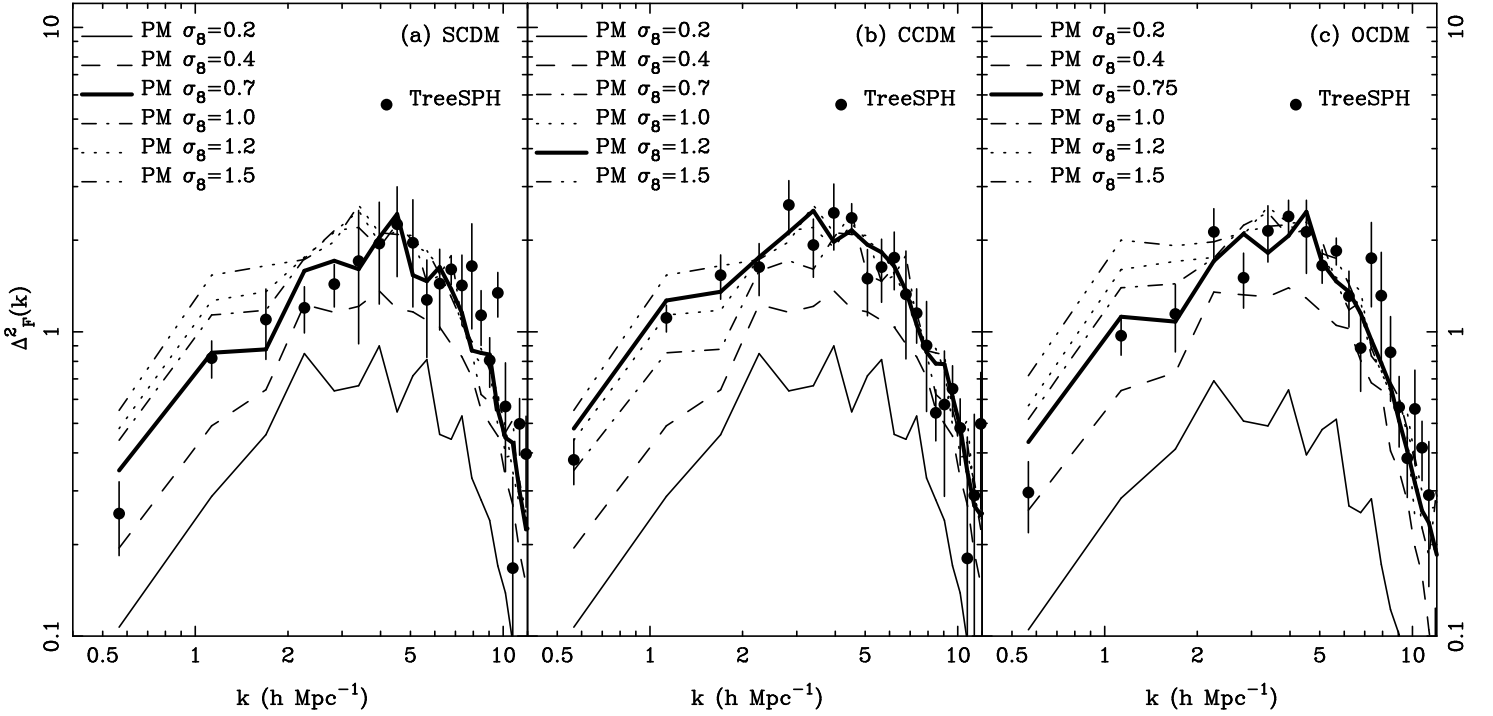}
\caption[junk]{\label{fluxpksig8}
Dependence of the flux power spectrum $\deltaf$ on the amplitude of
underlying mass fluctuations.  Results are shown at $z=3$ for the
(a) SCDM, (b) CCDM, and (c) OCDM models.
In each panel, points show $\deltaf$ for 100 spectra from the TreeSPH 
simulation, with error bars computed from the $1\sigma$ scatter among
five groups of 20 spectra each.
Lines show $\deltaf$ from 1000 spectra extracted from PM simulations
with the same initial fluctuations as the TreeSPH simulation but
different linear theory amplitudes, which are indicated in the legends.
In each case, the heavy solid line shows the PM results for the linear theory
amplitude used in the corresponding TreeSPH simulation.
}
\end{figure}

Figure~\ref{fluxpksig8} demonstrates the sensitivity of $\deltaf$
to the amplitude of linear theory mass fluctuations.
The points show $\deltaf$ measured from the
three TreeSPH simulations, with error bars calculated as in the previous
figures from the $1\sigma$ dispersion among five sets of 20 spectra each.
The lines show $\deltaf$ computed from PM simulations
that have the same initial
fluctuations as the corresponding TreeSPH simulations but different
linear theory amplitudes.  We evolve the PM simulations using the
appropriate cosmological parameters ($\Omega_0$, $\Lambda_0$, and $h$)
and extract spectra as described above, using the temperature--density
relation measured from the corresponding TreeSPH run
(e.g., $T_0=5600\;\K$, $\alpha=0.6$ for SCDM).
We label each curve by $\sigma_8$, the
rms linear theory mass fluctuation in $8\hmpc$ spheres
at $z=0$ implied by the adopted power spectrum normalization.
Figure~\ref{fluxpksig8} shows results at $z=3$, when the linear theory
amplitudes are lower by a factor of 4.0 in the two $\Omega=1$ models
and 2.71 in the open model.
We measure $\deltaf$ from 1000 lines of sight for each PM simulation,
enough that the fluctuations in the $\deltaf$ curves do not change
with the addition of more spectra.
Some fluctuations remain because we are always simulating a single
$11.111\hmpc$ box with the same initial phases and therefore have
a finite number of independent structures in the evolved gas distribution.
We use the same initial phases for the PM and TreeSPH runs in this
Section in order to limit the effect of this cosmic variance 
on the $\deltaf$ comparisons.

As expected, on large scales ($k \la 2\invhmpc$) the amplitude
of the flux power spectrum increases steadily with increasing
mass fluctuation amplitude.  On small scales the $\deltaf$ curves
turn over because of non-linear gravitational evolution and
blurring of the \lya\ spectra by peculiar velocities, and the curves
for $\sigma_8 > 0.4$ converge.
In each panel of Figure~\ref{fluxpksig8}, we use a heavy solid line
to indicate the $\deltaf$ curve from the PM simulation with the
correct linear theory amplitude.  In all cases, this curve provides
the best overall visual fit to the $\deltaf$ values obtained from the TreeSPH
simulation.
Figure~\ref{fluxpksig8} thus demonstrates two essential points:
the flux power spectrum at small $k$ is sensitive to the mass fluctuation
amplitude, making it an appropriate fitting constraint for 
$P(k)$ normalization, and the PM approximation is accurate enough 
for determining this normalization properly.

As mentioned above, we choose the value of $\obj$
for each PM fluctuation amplitude to match the mean flux decrement
of the TreeSPH spectra, $D_A(z=3)=0.36$.
As the mass fluctuation amplitude increases, more of the gas flows into 
near-saturated regions, and a higher value of $\obj$ is required to give
the correct $D_A$.
 For example, the PM simulations of SCDM
require a value of $\obj$ 2.4 times higher for
$\sigma_8=1.2$ than for $\sigma_{8}=0.7$, consistent
with the results from the TreeSPH simulations themselves
(see Croft \etal 1997a).
 
\begin{figure}[t]
\centering
\vspace{10.0cm}
\includegraphics{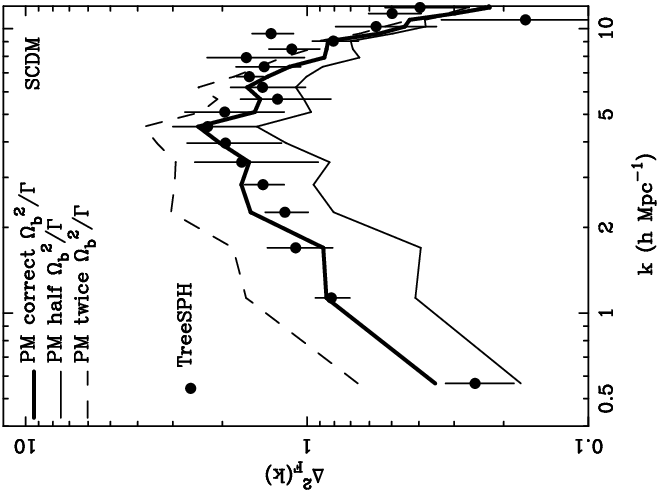}
\caption[junk]{\label{diffjfluxpk}
Dependence of $\deltaf$ on the value of $\obj$.
Filled circles (repeated from Fig.~\ref{fluxpksig8}a) show $\deltaf$
from the SCDM TreeSPH simulation.
The heavy solid line (also repeated from Fig.~\ref{fluxpksig8}a)
shows $\deltaf$ from the PM simulation with the correct mass
fluctuation amplitude and $\obj$ chosen to reproduce the mean
flux decrement of the TreeSPH spectra, $D_A=0.36$. 
The other lines show $\deltaf$ from the same PM simulation with 
$\obj$ increased (dashed line) and decreased (thin solid line)
by a factor of two, corresponding to lower and higher values of $D_A$,
respectively.  These results demonstrate the importance of
fixing $\obj$ to yield the observed $D_A$ when normalizing the
mass power spectrum on large scales.
}
\end{figure}

Figure~\ref{diffjfluxpk} highlights the importance of the 
mean flux decrement constraint to the normalization procedure.
The points and heavy solid line, repeated from Figure~\ref{fluxpksig8}a,
show, respectively, $\deltaf$ from the TreeSPH simulation and from
the PM simulation with the correct linear theory amplitude and
a value of $\obj$ that yields $D_A=0.36$.
The dashed line and thin solid line show $\deltaf$ from the same
PM simulation with the value of $\obj$ respectively increased and
decreased by a factor of two.  
These changes alter $\deltaf$ by about a factor of two at large scales,
roughly what one might expect from the physical
discussion in \S 2.1.  The changes to $\obj$ alter the constant $A$ in the 
$\tau-\rho$ relation~(\ref{eqn:tau}) by a factor of two, and this
constant serves as a sort of ``bias factor'' between density fluctuations
and optical depth fluctuations.  In studies of galaxy clustering,
the bias between galaxies and mass is not known a priori;
if it is assumed to be independent of scale, then one obtains the
shape of the mass power spectrum but not its amplitude.
We can obtain the shape {\it and amplitude} of the mass power spectrum
from \lya\ forest observations because the mean flux decrement
provides an observational constraint on the effective ``bias factor'' that is 
independent of the flux power spectrum itself --- 
$D_A$ measures the mean \lya\ opacity, while $\deltaf$ measures 
fluctuations about the mean. 
High precision estimates of the amplitude of mass fluctuations
require reliable measurements of $D_A$ from observations.
Current estimates of $D_A$ are not all 
consistent (e.g., compare Press et al.\ 1993 and
\cite{rauch97} with \cite{ZL93}), a situation that must be resolved
in order to make the most of our normalization procedure.

From equation~(\ref{eqn:tau}), it is clear that the value of
$\obj$ needed to match $D_A$ for a given density field will itself
depend on the adopted values of the IGM parameters $T_0$ and $\alpha$
and the cosmological parameters $\Omega_0$, $\Lambda_0$, and $h$,
which collectively determine $H(z)$.
To the extent that the approximation~(\ref{eqn:tau}) holds,
the effects of changing these parameters and 
of changing $\obj$ are degenerate
(except for the small influence of $\alpha$ on the
index of the $\tau-\rho$ relation).  Normalizing
to the observed $D_A$ therefore eliminates their
influence on the derived amplitude of mass fluctuations.
However, varying these parameters also changes
the amount of thermal broadening and peculiar velocity distortion
in the \lya\ spectrum (eq.~[\ref{eqn:tau}] ignores these effects),
and we must check that such variations do not thereby alter the inferred
$P(k)$ normalization.

\begin{figure}[t]
\centering
\vspace{10.0cm}
\includegraphics{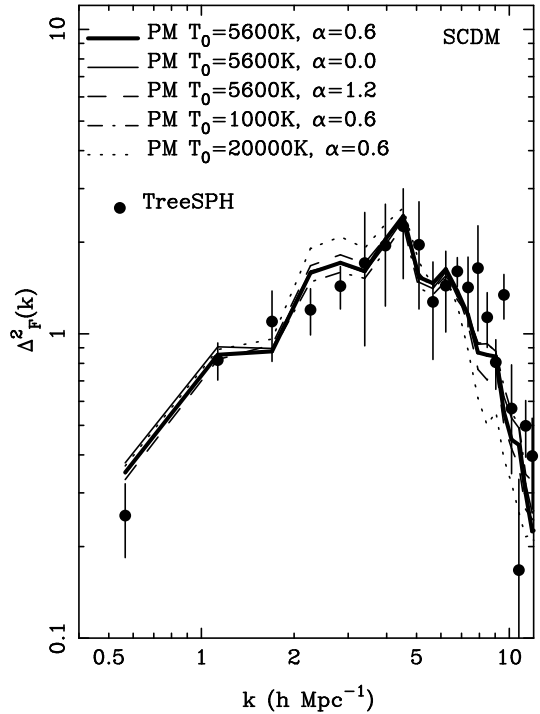}
\caption[junk]{\label{t0fluxpk}
Dependence of $\deltaf$ on the IGM parameters $T_0$ and $\alpha$.
The filled circles and the heavy solid line, repeated from
Fig.~\ref{fluxpksig8}a, show $\deltaf$ from the SCDM TreeSPH 
simulation and from the PM simulation with the same initial
conditions and temperature-density relation.
Other lines show $\deltaf$ from the PM simulation with different
values of $T_0$ and $\alpha$.  In each case,
$\obj$ is chosen to keep the mean flux decrement $D_A=0.36$.
With this constraint imposed, $\deltaf$ is insensitive to the
adopted values of $T_0$ and $\alpha$.
}
\end{figure}

Figure~\ref{t0fluxpk} illustrates the effects of $T_0$ and $\alpha$ on 
$\deltaf$ at fixed $D_A$.  The filled circles and heavy solid line,
repeated from Figures~\ref{fluxpksig8}a and~\ref{diffjfluxpk},
show $\deltaf$ obtained, respectively, from the SCDM TreeSPH simulation
and from the PM simulation with the same initial conditions
and the values $T_0=5600\;\K$, $\alpha=0.6$ that fit the
temperature--density relation of the TreeSPH run.
The other lines show $\deltaf$ obtained from the PM simulation
with different (rather extreme) values of $T_0$ and $\alpha$.
For each combination of $T_0$ and $\alpha$, we adjust $\obj$ so
that the mean flux decrement of the PM spectra is $D_A=0.36$;
for example, $\obj$ is 2.2 times higher for $T_0=20,000\;\K$
than for $T_0=5600\;\K$.  With the $D_A$ constraint imposed,
we see that $\deltaf$ is almost completely insensitive to the value of 
$\alpha$ and to a factor $\sim 4$ increase or decrease in $T_0$.
For $T_0=20,000\;\K$, the larger degree of thermal broadening
depresses $\deltaf$ on small scales, but even this extreme
variation does not significantly alter $\deltaf$ for $k \la 2 \invhmpc$.

Changes in the cosmic reionization history and the spectral shape of 
the UV background influence the low density IGM through the parameters
$T_0$ and $\alpha$ (\cite{hg97}).  
Since Figure~\ref{t0fluxpk} shows that changing $T_0$ and $\alpha$
does not alter $\deltaf$ if $D_A$ is held fixed, we conclude that
uncertainties in the reionization history and UV background spectral
shape do not affect our ability to normalize $P(k)$ accurately.

\begin{figure}[t]
\centering
\vspace{10.0cm}
\includegraphics{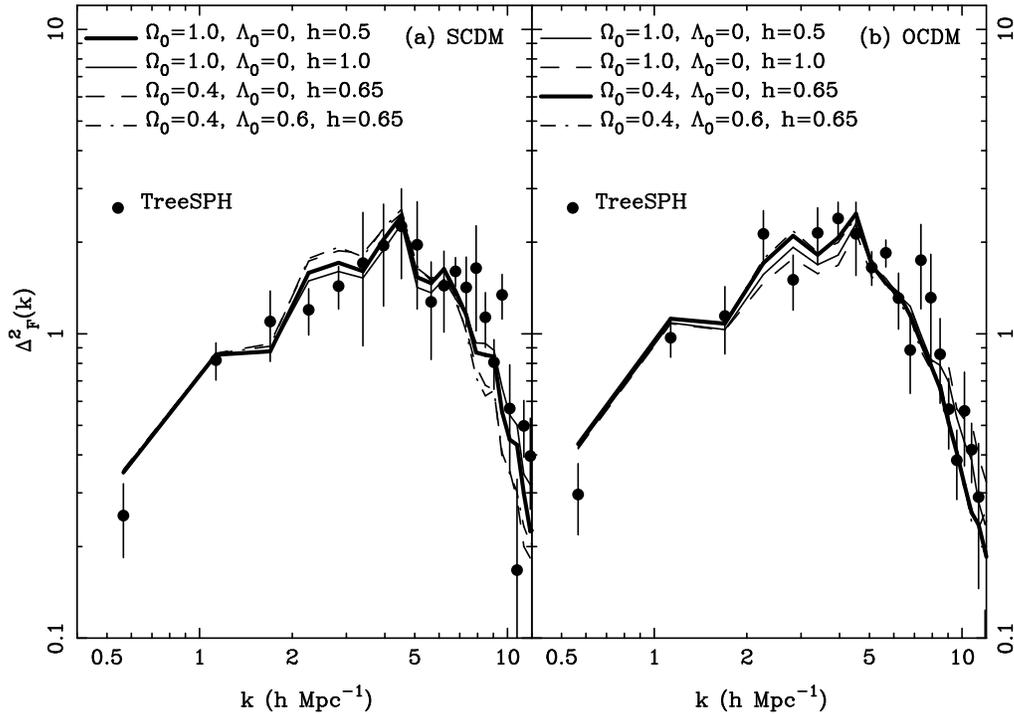}
\caption[junk]{\label{omegafluxpk}
Dependence of $\deltaf$ on cosmological parameters, with $D_A$ and the linear
theory mass fluctuations at $z=3$ held fixed.
Filled circles show $\deltaf$ from TreeSPH simulations of (a) the
SCDM model and (b) the OCDM model, as in Figs.~\ref{fluxpksig8}a
and~\ref{fluxpksig8}c.  Bold solid lines (also repeated from
Fig.~\ref{fluxpksig8}) show $\deltaf$ from
PM simulations with the same cosmological parameters.
Other lines show results from PM simulations evolved with different
cosmological parameters, indicated in the legends.
In each case, the amplitude of fluctuations at the start of the 
simulation ($z=15$) is chosen so that the linear theory amplitude
at $z=3$ is the same as that for the corresponding TreeSPH simulation.
}
\end{figure}

Figure~\ref{omegafluxpk} shows the influence of changing the cosmological 
parameters $\Omega_0$, $\Lambda_0$, and $h$ on $\deltaf$.
The filled circles in panels (a) and (b) show the TreeSPH results
for the SCDM and OCDM models, respectively, 
as in Figures~\ref{fluxpksig8}a and~\ref{fluxpksig8}c.
In each panel the heavy solid line (again repeated from 
the corresponding panel of Figure~\ref{fluxpksig8})
shows $\deltaf$ from the PM simulation with the correct 
mass fluctuation amplitude, temperature--density relation, 
and cosmological parameters.  The other lines show results
from PM simulations with the same linear theory density fluctuations
and temperature--density relation 
but different values of the cosmological parameters.
In each PM simulation we adopt a comoving box size of $11.111\hmpc$ and
scale the amplitude of the initial fluctuations (at $z=15$) 
so that their linear theory amplitude at $z=3$ matches that of the 
corresponding TreeSPH simulation.
We also adjust $\obj$ to keep $D_A=0.36$.

With $D_A$ and the linear theory mass fluctuations held fixed,
$\deltaf$ is insensitive to the adopted cosmological parameters.
Except in the cores of virialized objects, the non-linear mass density
field depends almost exclusively on the linear theory mass fluctuations,
independent of $\Omega_0$ and $\Lambda_0$ (\cite{wg90}; \cite{nusser97}).
Linear theory peculiar velocities are approximately proportional
to $\Omega^{0.6}$ (\cite{peebles80}), but at high redshift $\Omega$ is
always close to one; for $\Omega_0=0.4$, the $\Omega^{0.6}$ factor 
at $z=3$ is 0.83 in the open model and 0.99 in the flat, 
non-zero $\Lambda$ model.
The value of $H(z)$ depends on $\Omega_0$, $\Lambda_0$, and $h$,
but the overall scaling of optical depths with $H^{-1}(z)$ is removed
by the $D_A$ normalization.  For fixed $T_0$ and $\alpha$,
the importance of thermal motions relative
to Hubble flow and peculiar velocities is larger when $H(z)$ is smaller,
but even for small $H(z)$ the impact of thermal broadening on $\deltaf$
is insignificant.  
Other statistical properties of the flux might be able to detect
the direct influence of these cosmological parameters, but
our normalization of $P(k)$ will not depend on them.
Note, however, that once other observational constraints are imposed
(for example, the abundance of massive galaxy clusters at $z=0$, or
the amplitude of CMB anisotropies), the $P(k)$ predicted 
on Mpc scales at high redshift will depend on the adopted 
cosmological parameters.
Thus, the combination of $P(k)$ from QSO spectra
with other robust observational measures can provide critical
tests of cosmological models.  We will return to this issue in \S 5.

\begin{figure}[t]
\centering
\vspace{10.0cm}
\includegraphics{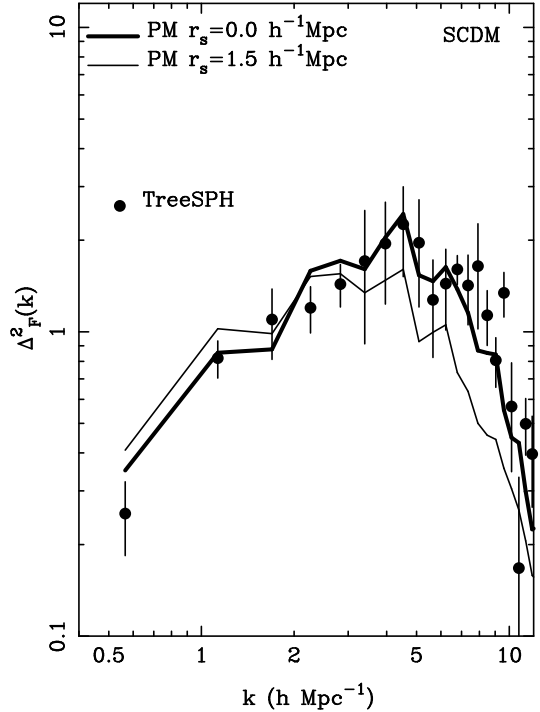}
\caption[junk]{\label{rsfluxpk}
Dependence of $\deltaf$ on small scale power in the initial fluctuations.
Filled circles and the heavy solid line show $\deltaf$ from
the TreeSPH and PM simulations of the SCDM model, as in Fig.~\ref{fluxpksig8}a.
The thin solid line shows $\deltaf$ from a PM simulation evolved
from the same initial conditions smoothed with a $1.5\hmpc$ Gaussian
filter, to approximate the loss of small scale power seen in
Fig.~\ref{pkscdm}.
}
\end{figure}

In our tests so far, we have evolved PM simulations from the same
initial fluctuations as the TreeSPH simulations, varying only the
amplitude.  However, Figure~\ref{pkscdm} shows that the mass power
spectrum inferred from the Gaussianized \lya\ flux is systematically
depressed on small scales, and we must check that this systematic
error in the {\it shape} of $P(k)$ at high $k$ does not change
the {\it amplitude} of $P(k)$ inferred by matching $\deltaf$
at lower $k$.  As shown in Figure~\ref{pkscdm}, the small scale
depression is roughly equivalent to smoothing the true $P(k)$
with a Gaussian filter of comoving radius $1.5\hmpc$.
Figure~\ref{rsfluxpk} shows the flux power spectrum
$\deltaf$ from the TreeSPH, SCDM simulation and from the PM
simulation evolved from the same initial conditions,
as in Figure~\ref{fluxpksig8}a.  The thin solid line shows 
$\deltaf$ from a PM simulation in which the initial fluctuations
are smoothed with a $1.5\hmpc$ Gaussian filter before they
are evolved forward.  As one might expect, $\deltaf$ at $z=3$
is somewhat lower at high $k$ because of the reduced small scale
power in the initial fluctuations.  However, on the larger
scales that we use for determining the $P(k)$ normalization,
the smoothing of the initial conditions has little effect.
One could slightly improve the accuracy of the normalization 
procedure by amplifying the small scale power in the Gaussianized
flux $P(k)$ before the PM evolution step, but to keep
the presentation of our method reasonably simple, we will not
include such corrections in this paper.

The results of this Section are encouraging and easy to summarize.
The flux power spectrum $\deltaf$ depends directly on the
linear theory amplitude of mass fluctuations, and it is insensitive
to the adopted values of IGM parameters and cosmological parameters
as long as one adjusts $\obj$  to match the mean flux decrement
$D_A$ to the observed value.  The PM approximation provides an
accurate and inexpensive method to compute $\deltaf$.
We therefore have a practical and robust procedure for determining
the normalization of $P(k)$, once its shape has been derived
from the Gaussianized \lya\ flux as described in \S 2.2.

\subsection{Effects of noise, resolution, and continuum fitting}
Real QSO spectra may have coarser resolution than our simulated spectra, 
and they are affected by photon noise from the QSO and the sky
and by instrumental readout noise.  The throughput of the atmosphere
and instrument as a function of wavelength may not be known exactly,
and in any case one does not have precise a priori knowledge of the
underlying QSO continuum in the \lya\ forest region. Therefore, the flux
level corresponding to zero absorption is usually determined from
the spectrum itself by a local continuum fitting procedure.
In this Section, we will add noise to the TreeSPH spectra, degrade
their spectral resolution, and apply local continuum fitting
to see how these observational realities affect our ability to 
determine $P(k)$.

To model the effect of limited spectral resolution, we take the simple
approach of rebinning the TreeSPH spectra (which have 
$\sim 2\kms$ pixels) into larger pixels --- a ``top hat'' smoothing
of the transmitted flux.
We then add photon noise with a specified signal-to-noise
ratio S/N in the continuum.
For each pixel, we draw the photon noise from a Poisson distribution
with a mean proportional to the transmitted flux level,
effectively assuming the limit where noise from the QSO itself
dominates over noise from the sky.  A more thorough treatment
of sky photon noise requires specifying the QSO and sky flux
at the observed \lya\ forest wavelengths; we defer detailed tests
of these effects to a later paper where we analyze a large
observational data set.  In addition to photon noise, we add
noise drawn from a Gaussian distribution with zero mean and standard deviation
$\Delta F=0.01$ (where $F=1$ represents the unabsorbed continuum),
independent of the pixel flux level, to model instrument
readout noise.  

We try three different combinations of spectral resolution/noise parameters:\\
\noindent(1) High resolution (4 $\kms$ pixels), high S/N (50 per resolution
element in the continuum). These values are characteristic of
a typical Keck HIRES spectrum (e.g., \cite{hu95}).  \\
\noindent(2) High resolution (4 $\kms$ pixels), low S/N (10 per
resolution element).  \\
\noindent(3) Low resolution (40 $\kms$ pixels), low S/N (10 per 
resolution element).\\
\noindent
Since we degrade the spectral resolution by top hat rebinning,
a ``resolution element'' is simply a pixel of the rebinned spectrum.

The implicit assumptions in standard continuum fitting procedures 
are that the QSO continuum and instrumental response vary slowly as
a function of wavelength and that the highest observed flux levels
correspond to the unabsorbed continuum (plus noise).
In high resolution, echelle spectra, the continuum is often determined
separately for each echelle order by (1) fitting a 3rd-order polynomial 
to the data points, (2) rejecting all points that lie more than 2-$\sigma$ 
below this polynomial fit, and 
(3) repeating the process with the points that remain,
iterating until the continuum fit converges.  
This is the procedure that we will use here 
(it was also used by \cite{dhwk97}).
Each of our simulated spectra is periodic on the scale of the
simulation box, which at $z=3$ corresponds to $36\AA$ in the SCDM and CCDM
simulations and $27\AA$ in the OCDM simulation.
This scale is shorter than the $\sim 45\AA$ length of a typical Keck 
HIRES echelle order.
In order to partially account for this difference in length and
reduce edge effects, we periodically extend each spectrum by an equal
length on either side of the simulation box 
to create a $45\AA$ region. We fit the continuum to this region.
Nonetheless, the systematic depression of the continuum below its
true value should be larger in our simulated observations than
in real data because there is less chance of finding a region
of genuinely low absorption in the short spectra we have available.
Our tests should therefore give an upper limit to the effects
of continuum fitting on the determination of $P(k)$.

\begin{figure}[t]
\centering
\vspace{10.5cm}
\includegraphics{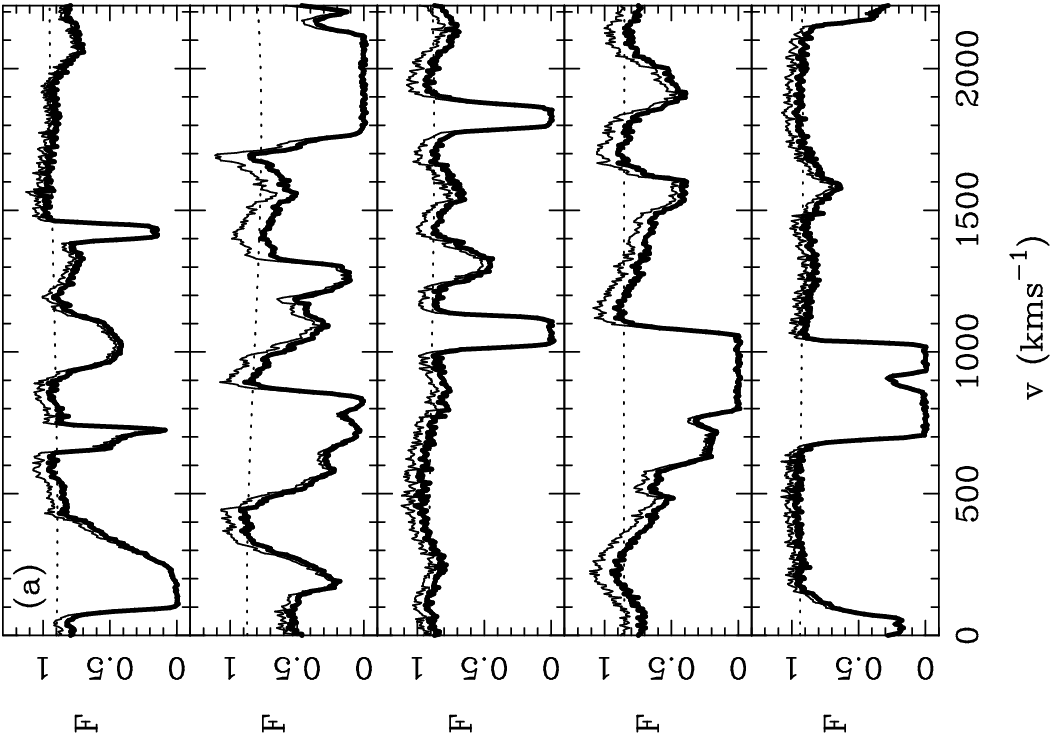}
\includegraphics{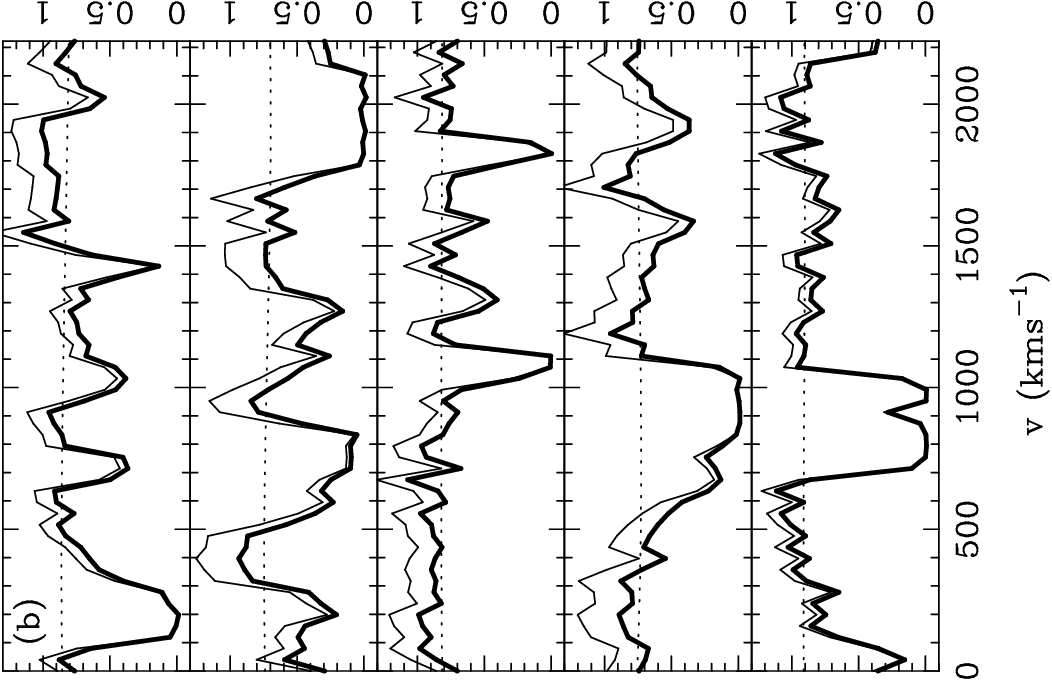}
\caption[junk]{\label{cf}
The effects of continuum fitting and renormalization on simulated
spectra from the SCDM model at $z=3$, for (a) $4\;\kms$ pixels
and S/N=50, (b) $40\;\kms$ pixels and S/N=10.
In each panel, the heavy solid lines shows the simulated spectrum
with degraded spectral resolution and added noise.
The dotted line shows the continuum estimated by the 3rd-order
polynomial fitting procedure described in the text.
The thin solid line shows the renormalized spectrum, with the
original flux in each pixel divided by the fitted continuum level.
}
\end{figure}

Figure~\ref{cf} illustrates the effect of continuum fitting
on several example spectra from the SCDM simulation,
for the high resolution, high S/N and low resolution, low S/N
observational parameters.
In each panel, the heavy solid line shows the TreeSPH spectrum,
rebinned to the appropriate spectral resolution, with added noise.  
The dotted line shows the fitted continuum.
Light solid lines show the renormalized spectra, with the original
flux in each pixel divided by the local continuum level.
Clearly the main effect of continuum fitting and renormalization is to 
remove absorption (boost the flux) in low density regions, 
since the fitted continuum is systematically depressed there.
The effect is largest for the low resolution, low S/N spectra, 
where the mean flux decrement is reduced by $\sim 14\%$.
In the high resolution, high S/N spectra, continuum renormalization
reduces the mean flux decrement by $\sim 5 \%$.

\begin{figure}[t]
\centering
\vspace{10.0cm}
\includegraphics{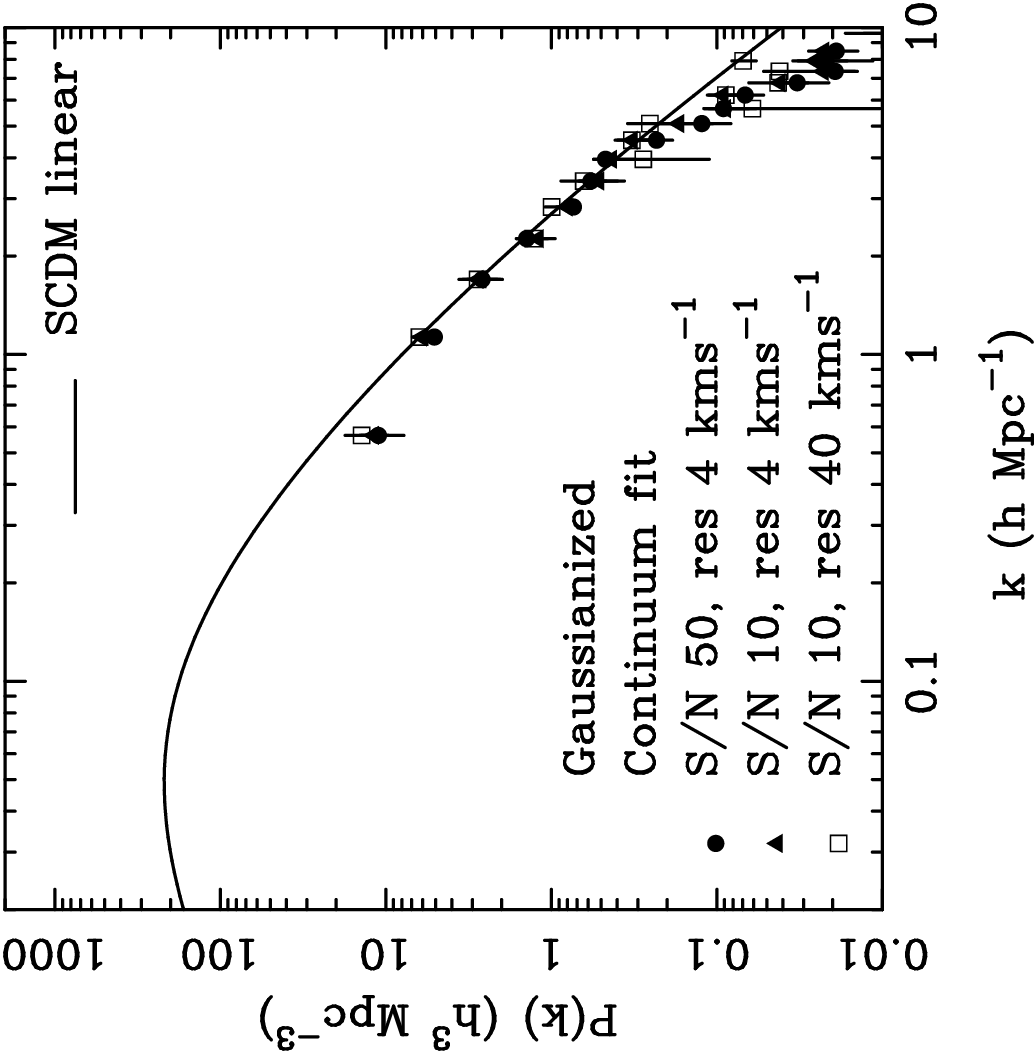}
\caption[junk]{\label{pkcf}
The effects of noise, spectral resolution, and continuum fitting on the
shape of the power spectrum derived from the Gaussianized flux.
Points show $P(k)$ estimated from the continuum-renormalized,
degraded TreeSPH spectra, with the observational parameters
indicated in the legend.  Error bars show the $1\sigma$ dispersion
among five sets of 20 spectra each.  
Amplitudes of the derived $P(k)$ are chosen to match
the SCDM linear theory power spectrum, shown by the solid line.
These results can be compared to those from noiseless spectra,
shown in Fig.~\ref{pkscdm}.
}
\end{figure}

Figure~\ref{pkcf} shows how noise and continuum fitting influence
the shape of $P(k)$ derived from the Gaussianized flux.
These results from the degraded spectra can be compared to those
from the noiseless spectra shown in Figure~\ref{pkscdm}.
The low resolution, low S/N spectra give somewhat noisier $P(k)$
estimates at small scales, but noise and continuum fitting do
not appear to distort the shape of the inferred $P(k)$.
The largest scale points ($2\pi/k=11.111\hmpc$) lie a factor $\sim 2$
below the linear theory power spectrum.  
Since there are only a few modes at this scale in our simulation box,
this depression of large scale power could be a statistical fluctuation,
but it could also be a systematic effect of continuum fitting, which
tends to even out large scale fluctuations.
The uncertainties in continuum determination may ultimately
set the upper limit to the scale where $P(k)$ can be determined
reliably from the \lya\ forest. This question is best addressed
in the context of specific observational data using larger volume
simulations, and  we therefore defer a more thorough investigation of
continuum fitting procedures and their effects to future work.
For now, we note that Figure~\ref{pkcf} shows that there is an
interesting range of scales over which noise and continuum fitting
do not pose serious problems for the determination of the shape of $P(k)$.

\begin{figure}[t]
\centering
\vspace{10.0cm}
\includegraphics{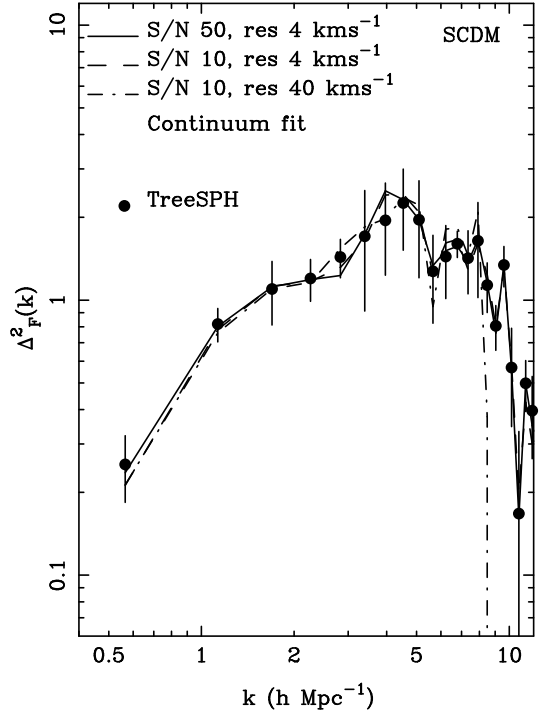}
\caption[junk]{\label{fluxpkcf}
The effects of noise, spectral resolution, and continuum fitting on
the flux power spectrum.
Points with error bars (repeated from Fig.~\ref{fluxpksig8}a)
show $\deltaf$ from noiseless, TreeSPH, SCDM spectra.
Lines show $\deltaf$ measured from continuum--renormalized
spectra with the noise and resolution parameters indicated in the legend.
}
\end{figure}

Noise and continuum fitting have even less impact on the
normalization of $P(k)$ using $\deltaf$.
Figure~\ref{fluxpkcf} shows $\deltaf$ in the SCDM model measured from
the noiseless TreeSPH spectra (points with error bars, repeated
from Figure~\ref{fluxpksig8}a) and from the degraded,
continuum--renormalized spectra.  The flux power spectrum cuts off
on small scales in the low resolution spectra as one might expect,
but $\deltaf$ is virtually unchanged on the larger scales that
we use for normalization.  Even with $40\;\kms$ pixels the suppression
of power is limited to scales a factor of two smaller than those
where the Gaussianized flux power spectrum matches the linear theory
mass $P(k)$ (see Figure~\ref{pkcf}), suggesting that we could use
still coarser resolution spectra for power spectrum determination without 
losing useful information.  (However, continuum determination
might be more problematic in lower resolution spectra, so we might 
ultimately lose information on {\it large} scales.)
In general, pixel--scale variations in detector response (e.g.,
flat--fielding errors, cosmic rays) only affect $\deltaf$ and $P(k)$ at
high $k$, where our recovery is limited in any case by the effects of 
peculiar velocities, thermal motions, and non-linear evolution.
It is only large scale, coherent variations that are cause for concern.
As Figure~\ref{fluxpksig8} shows, the fluctuation amplitudes 
(per logarithmic $k$ bin) predicted by CDM models are of order 
unity on the scales considered in this paper, so any systematic 
errors would have to be quite substantial  to influence
the recovered power spectrum.

\section{A test of the entire procedure on simulated observations}
We have now tested all of the pieces of our power spectrum recovery
method individually.  For the tests in \S 2, we have always used
the same initial density field that was used in the TreeSPH simulations,
in order to separate errors in the recovery procedure from the
statistical fluctuations in finite--volume realizations of a given model.
In this Section we will treat spectra from the TreeSPH simulations
as if they were observational data, and attempt to recover the mass
power spectrum from them without using prior knowledge of the initial
conditions or cosmological parameters.  Comparison of the results
to the models' true linear power spectra will provide end--to--end
tests of the full $P(k)$ recovery procedure.

As our starting point, we take the low resolution, low S/N spectra from \S 2.4.
These are the worst case observational parameters that we have considered,
and if the procedure works well for these spectra then it should be
very widely applicable in practice.
We generate 100 spectra for each of the three cosmological models,
fitting a continuum to each spectrum as described in \S 2.4.
We will again compute error bars from the dispersion among five
sets of 20 spectra, each set roughly equal in redshift path 
length to a single observed QSO spectrum.
One caveat to bear in mind is that these five simulated ``QSOs'' probe
the structure in a single simulation box; on scales comparable to
the box, there are only a handful of Fourier modes, and their average
power may not equal the true cosmic average predicted by the model.
Real spectra of the same total path length would sample
more independent large scale fluctuations and should therefore
yield an average $P(k)$ closer to the true value.
However, the dispersion in $P(k)$ estimates from one spectrum to another
should also be larger if they are fully independent, so the
statistical error bars that we derive will be somewhat underestimated.
 
To recap, the steps of the $P(k)$ recovery procedure are as follows:\\
\noindent(1) Gaussianization. We reassign pixel flux values so that
they have a Gaussian PDF but the same rank order as in the original spectrum.
This step yields an unnormalized estimate of the linear density contrast
field along each line of sight.\\
\noindent(2) Measurement of the shape of $P(k)$. We estimate the 
1-dimensional power spectrum of the Gaussianized flux using an FFT.
We convert to the 3-dimensional $P(k)$ using equation~(\ref{eqn:invert}).\\
\noindent(3) Determination of the amplitude of $P(k)$.\\
\noindent(a) We use $P(k)$ from step (2) to create 
realizations of an initial, random phase, linear density field.
We evolve each density field forward under gravity
using a PM code. We choose the cosmological parameters 
relevant to the SCDM model ($\Omega_{0}=1$, $\Lambda_{0}=0$, $h=0.5$) 
to do this, but we have seen in \S 2.3
that any other choice will make essentially no difference to the results.\\
\noindent(b) At each of several output times,
we generate spectra using a power law
temperature--density relation (any reasonable choice of $T_0$ and $\alpha$
in eq.~[\ref{eqn:td}] will do) and the value of $\obj$ that yields
the same mean flux decrement as the input spectra, in this case $D_A=0.36$.
Because the linear density contrast grows in proportion to the expansion
factor in an $\Omega=1$ universe, the multiple output times of a single
simulation are physically equivalent to the outputs at a fixed time (or
redshift) of simulations with different amplitudes of the linear theory mass
power spectrum.\\
\noindent(c) We estimate $\deltaf$ for each mass fluctuation amplitude, 
averaging results obtained from several realizations of the initial 
density field that have the same $P(k)$ but different random phases. 
We compare these $\deltaf$ values to those measured from the ``observed''
spectra, on large scales.  The normalization of $P(k)$ is found 
by linearly interpolating between the output times (fluctuation amplitudes)
that reproduce the $\deltaf$ amplitude most closely.

\begin{figure}[t]
\centering
\vspace{10.0cm}
\includegraphics{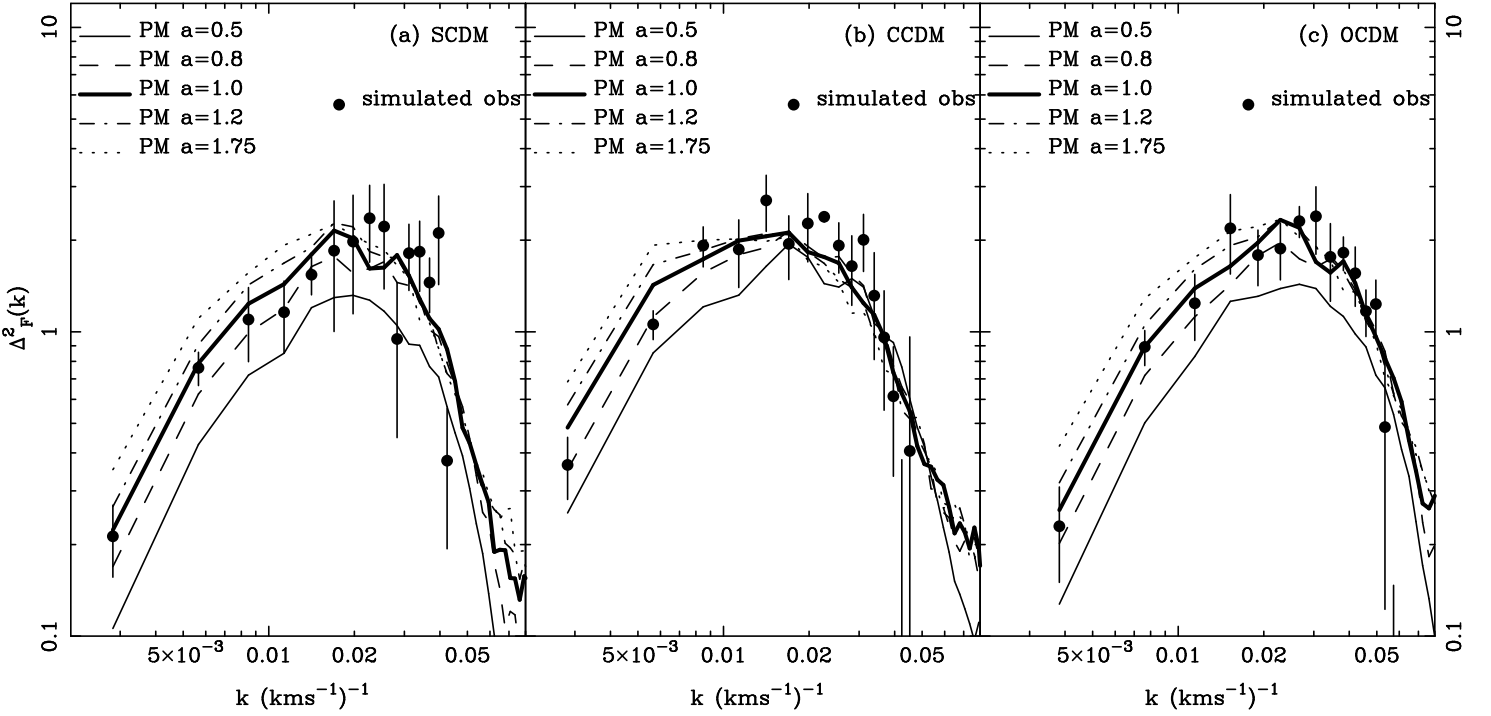}
\caption[junk]{\label{simobsfluxpk}
Normalization of the recovered $P(k)$ in tests on simulated observations
from the (a) SCDM, (b) CCDM, and (c) OCDM models.
In each panel, points show $\deltaf$ from simulated, continuum--renormalized
spectra with S/N=10 and $40\;\kms$ pixels.
Error bars show the $1\sigma$ dispersion among five sets of 20 spectra.
The heavy solid line shows the average $\deltaf$ from four PM
simulations with the $P(k)$ obtained from the Gaussianized flux of
the simulated data and the same linear theory fluctuation amplitude
as the corresponding TreeSPH simulation, i.e., the ``correct'' normalization.
Other lines show $\deltaf$ obtained from earlier and later outputs
of the PM simulations, corresponding to different linear theory
amplitudes proportional to the expansion factor $a$.
}
\end{figure}

We apply this procedure to the three CDM models in turn.
For the normalization step, we use four different random phase 
realizations in each case, with PM simulation boxes $11.111 \hmpc$
on a side, matched to the TreeSPH simulations.
We extract spectra along 1000 lines of sight from each simulation cube
at each output.
Figure~\ref{simobsfluxpk} shows the $\deltaf$ results used in 
the normalization step for the three models.
For the wavenumber $k$, we now use $(\kms)^{-1}$ units,
since these can be related directly to observed wavelength.
The conversion from $\kms$ to comoving $\hmpc$ would depend on
our adopted cosmological parameters.  At $z=3$, the SCDM and CCDM simulation
cubes are $2222\;\kms$ on a side, but the OCDM simulation
is $1647\;\kms$ on a side. Therefore, we measure 
$\deltaf$ over a different range of $k$ values in the open model.

The heavy solid line in each panel of Figure~\ref{simobsfluxpk}
shows results from the PM output with the same linear theory
mass fluctuation amplitude as the corresponding TreeSPH simulation.
Other lines show $\deltaf$ for output times with smaller and larger
expansion factors $a$.
We adopt $\Omega=1$ in the PM simulations, so the amplitude
of the linear density fluctuations scales with $a$.
On large scales, the $\deltaf$ values from the simulated observations
are usually matched by a PM output within 
$\Delta a=0.2$ of the correct value, indicating that the normalization 
procedure recovers the correct amplitude to an accuracy of $\sim 20\%$
or better.
To obtain the final normalization for $P(k)$, we take the four 
$\deltaf$ points with $k < 0.015 \;\invkms$ and use linear
interpolation among the PM outputs to find the amplitude that
best fits that point.  We average the four results to obtain
the $P(k)$ amplitude and take the scatter between them as an
estimate of the normalization error.
Analysis  of  larger volume simulations or a large sample of observational
results would warrant a more sophisticated treatment of the 
amplitude fitting and normalization error.

\begin{figure}[t]
\centering
\vspace{10.0cm}
\includegraphics{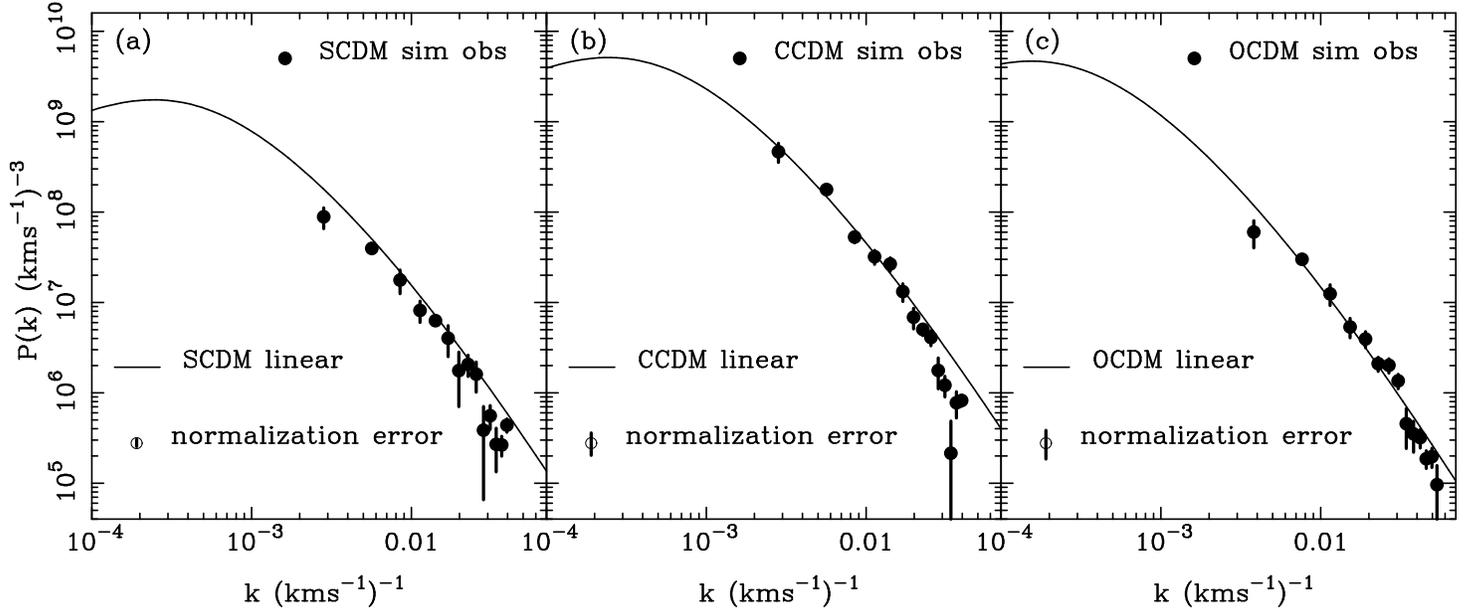}
\caption[junk]{\label{simobspk}
Tests of the power spectrum recovery on simulated observations
created from the (a) SCDM, (b) CCDM, and (c) OCDM simulations.
In each panel, solid curves show the true linear theory mass 
power spectrum at $z=3$.  Points show $P(k)$ recovered from
the Gaussianized flux of the simulated spectra, normalized by
matching the flux power spectrum as shown in Fig.~\ref{simobsfluxpk}.
Error bars on individual points show the $1\sigma$ dispersion
among five sets of 20 spectra each.
The uncertainty in the overall amplitude resulting from the $\deltaf$
fitting is indicated by the ``normalization error'' in the lower
left corner of each panel.
}
\end{figure}

Figure~\ref{simobspk} shows the final product, 
the normalized estimates of $P(k)$ for the three models.
We again plot the results in the directly observable length units, $\kms$.
Since $P(k)$ has dimensions of length$^{-3}$, the scaling 
between $\hmpc$ and $\kms$ affects both the $x$ and $y$ axes.
Error bars attached to the individual points represent the
$1\sigma$ dispersion in $P(k)$ obtained from the five sets
of 20 spectra.  The error bar in the lower left corner of each
panel indicates the uncertainty in the overall amplitude derived from
the normalization procedure; the full set of $P(k)$ points can
be coherently shifted up or down by this amount.
Solid curves in Figure~\ref{simobspk} show the true linear theory
power spectra of the three cosmological models.

Figure~\ref{simobspk} is the principal result of this paper.
It demonstrates that the procedure we have described can
recover the correct shape and amplitude of the linear theory
mass power spectrum at high redshift, even when it is applied to 
continuum--fitted spectra with a relatively low signal-to-noise ratio
and only moderate spectral resolution.

\section{Application to the spectrum of Q1422+231}

In this Section we present an illustrative application of our $P(k)$
recovery method to a single QSO spectrum. We will analyze larger
data sets and carry out more detailed comparisons to the predictions
of cosmological models in future work.

We analyze a Keck HIRES spectrum of the QSO Q1422+231 ($z_{\rm QSO}=3.63$) 
kindly provided by A. Songaila and L. Cowie. 
Other analyses of the \lya\ forest and associated CIV forest of
this QSO spectrum appear in Songaila \& Cowie (1996) and Kim et al.\ (1997).
The region of the
spectrum between \lya\ and Ly$\beta$ covers a redshift range 
$\Delta z\sim 0.7$, over which evolution of the universe has
a non-negligible impact. The strongest effect is the changing
relation between optical depth and mass overdensity
caused by the expansion of the universe and the change in the
Hubble expansion rate.  For photoionized gas expanding with the
Hubble flow in an $\Omega=1$ universe, the \lya\ optical depth
evolves as $(1+z)^{4.5}$ at constant UV background intensity.
Following Rauch \etal (1997), we scale the optical depths 
in the spectrum using this
relation to the values they would have at the central redshift of
absorption ($z_{\rm abs}=3.2$). The mean flux decrement that we
obtain for this spectrum is $D_A=0.37$, slightly below the
value $D_A=0.41$ implied by the results of Press \etal (1993).
In order to minimize the effect of the evolution in spatial
scales, we scale all pixels to the size they would have in $\kms$ at
$z_{\rm abs}$.  The maximum resulting change in pixel size is $\sim 10\%$.
In practice, this rescaling means that the pixels, which have a constant
separation in observed wavelength, are treated as having a constant
separation in velocity, $\Delta v = 4\;\kms$.
This scaling removes the first-order effect of the expansion of the 
universe over $\Delta z$.  Second-order effects due to the change in 
the Hubble parameter caused by deceleration will remain, but 
these should be small.

Once we have carried out the above rescalings, we Gaussianize the spectrum and 
calculate $P(k)$ as described in \S 2.2. 
Unlike the simulated spectra, the real spectrum does not have
periodic boundaries, so that the 1-dimensional power spectrum we measure 
by using an FFT is a convolution of the true 1-dimensional power spectrum 
with a top hat window function that represents the finite length 
of the QSO spectrum.  Because the QSO spectrum is much 
longer than the largest scale on which we will estimate $P(k)$,
the effect of this convolution is negligible.
Since we have only a single spectrum, we cannot calculate error bars
on $P(k)$ as we did with the ensemble of simulated observations.
Instead we split the $k$ range of interest into logarithmically
spaced bins, estimate $P(k)$ itself by averaging over the Fourier
modes in each bin, and estimate the uncertainty in $P(k)$ from the
Poisson errors based on the number of discrete $k$ modes in the bin
(i.e.,  $\Delta P(k) = P(k)/\sqrt{N}$, where $N$ is the number of
Fourier modes in the bin).

\begin{figure}[t]
\centering
\vspace{10.0cm}
\includegraphics{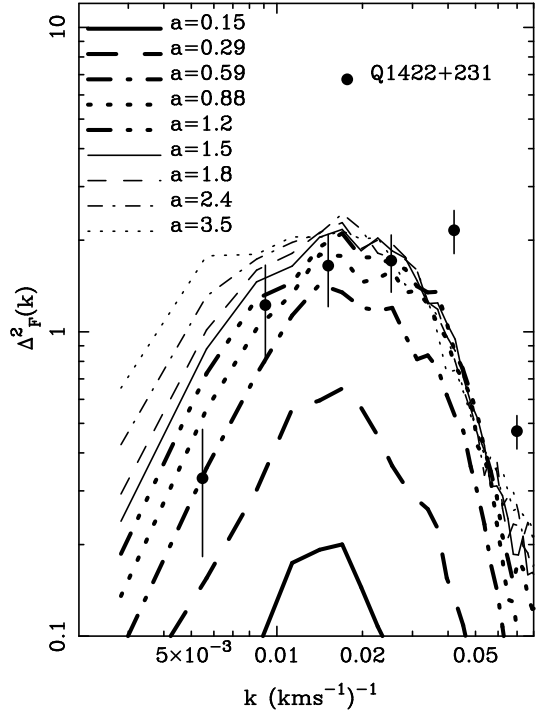}
\caption[junk]{\label{fluxpkq1422}
Normalization of the mass power spectrum derived from the \lya\ forest
of Q1422+231.
Filled circles show the flux power spectrum for Q1422+231,
with error bars computed from Poisson errors in the number of
Fourier modes in each logarithmic $k$ bin.
Lines show the average flux power spectra from four PM simulations evolved
from Gaussian initial conditions with the $P(k)$ shape estimated
from this QSO spectrum, at various expansion factors $a$ corresponding
to different linear theory mass fluctuation amplitudes.
For each of the four lowest $k$ points, we estimate the linear
theory amplitude required to match the observed $\deltaf$ by
linear interpolation between the two bracketing outputs.
The $P(k)$ normalization and its uncertainty are determined from
the mean and standard deviation of these four estimates.
}
\end{figure}

We use the estimated $P(k)$ to set up the initial density field for 
the PM code, as described in \S 2.3.
We linearly interpolate between the binned estimates of $P(k)$ to assign
power to all $k$ modes in the initial conditions.
As in our simulation tests from \S 3, we generate four 
different random phase realizations of this initial power spectrum,
adopting $\Omega=1$ and a comoving simulation box size $11.111\hmpc$
($2222\;\kms$ at $z=3$).
We evolve the simulations forward so that the 
scale factor $a$ grows by a factor of 24 in 120 steps of equal $\Delta a$.
We create artificial spectra with $D_A=0.37$ at several different output times
and measure the flux power spectrum.
The resulting values of $\deltaf$ are shown in Figure~\ref{fluxpkq1422}, 
together with $\deltaf$ for Q1422+231.
In  Figure~\ref{fluxpkq1422} we label the different simulation
output times with expansion factors $a$ relative to the expansion 
factor $a=1.0$ at which the simulations reproduce the observed $\deltaf$.
As in \S 3, we find the normalization of $P(k)$ using the four largest
scale points, estimating the amplitude required to match each point
by linearly interpolating between the two outputs that bracket the point.
We adopt the standard deviation of the four estimates as our estimate
of the overall normalization uncertainty. We find that 
the $1 \sigma$ error on the amplitude $a$
is $16 \%$. This error does not include the sampling variance that
 results from the use of only one QSO spectrum, and which could be
of comparable magnitude, or even greater.
Another source of uncertainty is the value of $D_A$ used in
the normalizing simulations.
If we had adopted the Press \etal (1993) value of $D_A$ instead 
of the mean decrement measured from Q1422+231, then our estimated
mass fluctuation amplitude would be lower by $\sim 10\%$
(20\% in $P(k)$). On the other hand, as the Zuo \& Lu (1993) estimate of 
 $D_A$ is somewhat lower, assuming their value instead 
would yield a higher amplitude for $P(k)$.

\begin{figure}[t]
\centering
\vspace{10.0cm}
\includegraphics{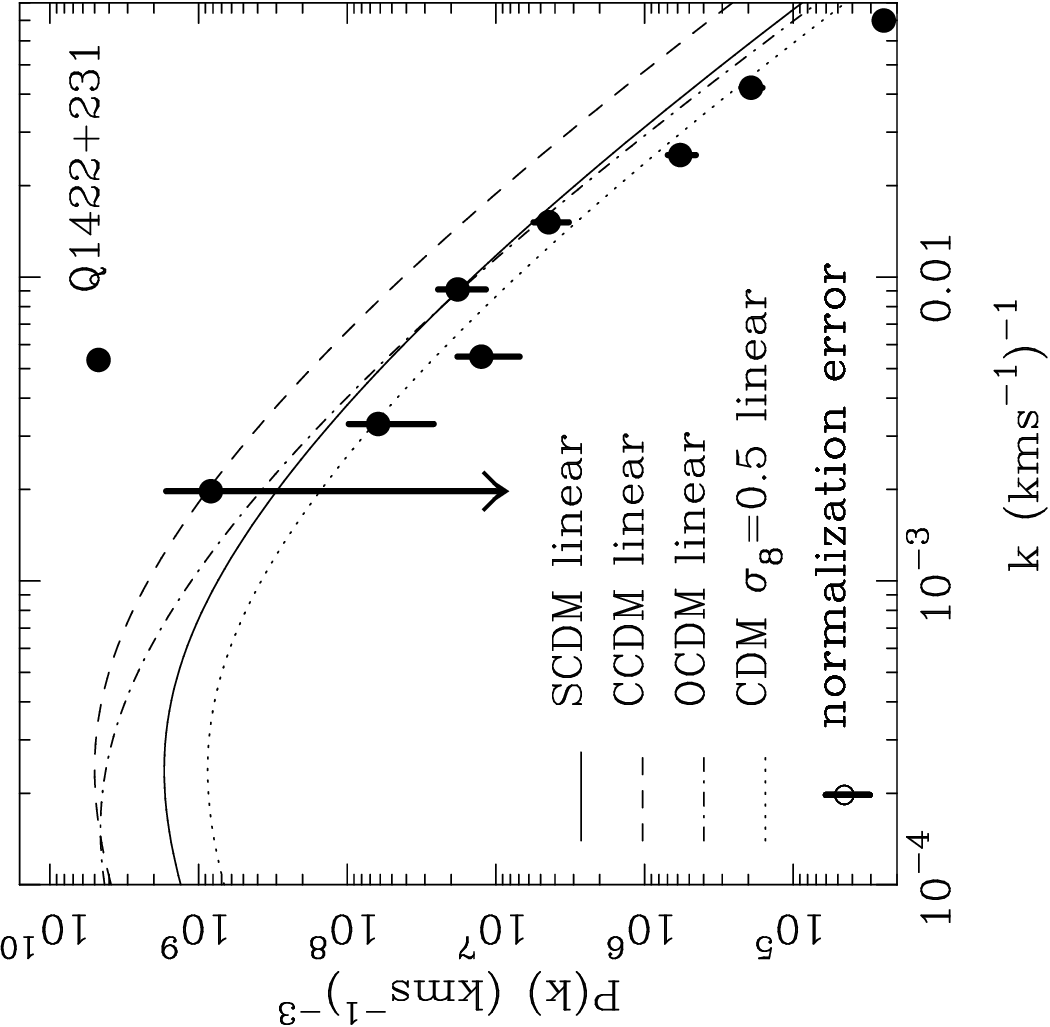}
\caption[junk]{\label{pkq1422}
The linear mass power spectrum $P(k)$ recovered from the \lya\ forest 
of Q1422+231 (filled circles).  Error bars on individual points are
computed from the Poisson errors in the number of Fourier modes in
each logarithmic $k$ bin.  The additional uncertainty in the $P(k)$ 
normalization, which affects all points coherently, is shown in
the lower left corner. Curves show the linear theory power spectra
at $z=3.2$, corresponding to the central absorption
 wavelength of Q1422+231, for the
three CDM models discussed earlier, and for an $\Omega=1$, $h=0.5$
CDM model with a lower fluctuation amplitude, corresponding to
$\sigma_8=0.5$ at $z=0$.
}
\end{figure}

Figure~\ref{pkq1422} shows the normalized $P(k)$ derived from 
the spectrum of Q1422+231.  The ``normalization error'' in the
lower left corner indicates the uncertainty in the overall amplitude
from the normalization procedure, and error bars on individual
points are based on Poisson errors in the number of Fourier modes
in each $k$ bin.  Note that the tests shown in Figure~\ref{simobspk}
imply that $P(k)$ may be systematically underestimated on the smallest scales,
$k \ga 0.025\;\invkms$.  

The curves in Figure~\ref{pkq1422} show the linear theory power
spectra of the SCDM, CCDM, and OCDM models at $z=3.2$.
The shape of the derived $P(k)$ appears to be broadly compatible with 
the shape predicted for CDM, roughly an $n=-2$ power law on the scales
probed by this measurement.
The best fit amplitude seems somewhat lower than that in any of our 
three models, and this difference is
seen at large scales where the loss of small-scale power in our tests
 is negligible.  For comparison, we also show the linear theory power
spectrum for an $\Omega=1$, $h=0.5$ CDM model with a lower amplitude,
$\sigma_8=0.5$.  This model fits the data fairly
well over the range of $k$ where the $P(k)$ recovery is reliable.
Since Figure~\ref{pkq1422} is based on a single QSO spectrum, it 
would be premature to draw strong conclusions about which cosmological
models are compatible or incompatible with the data, but it is
encouraging that the results derived by our method are roughly in
line with theoretical expectations.
Comparison to Figure~\ref{pklinen-1} shows that the large scale
power detected in the spectrum of Q1422+231 is incompatible with
a model in which the \lya\ forest is produced by a superposition
of unclustered, Voigt-profile lines.

\section{Summary and Discussion}

We have presented a method for recovering the linear
power spectrum of mass fluctuations from QSO \lya\ forest spectra.
The method is motivated by a simple, approximate
description of the relation between \lya\ optical depth and
underlying mass density in the ``fluctuating IGM'' scenario
for the origin of the \lya\ forest.  
We have carried out extensive tests on artificial QSO spectra
created from realistic hydrodynamic cosmological simulations
to show that the method successfully recovers both the 
shape and the amplitude of the mass power spectrum on an
interesting range of scales.
A preliminary application to the spectrum of Q1422+231 gives results
compatible with a low amplitude CDM model. 

On small scales, roughly $2\pi/k < 1.5\hmpc$ comoving,
our method fails to recover the initial $P(k)$ because of
non-linear gravitational evolution and the effective smoothing of
the \lya\ spectrum produced by peculiar velocities and thermal broadening.
Our present investigation does not tell us the largest scale out to
which our method can recover $P(k)$; it appears to work from
$2\pi/k \sim 1.5\hmpc$ up to the $11.111\hmpc$ scale of our
simulation boxes.  There is a hint from the tests in \S 2.4
that local continuum fitting may artificially suppress power
at the largest of these scales, but because our simulations
contain only a few modes at this wavelength, it is difficult
to tell whether this is a genuine effect or a statistical fluctuation.
The continuum--fitting issue requires further investigation with
larger volume simulations.  For purposes of measuring $P(k)$,
techniques that fit a low-order continuum over the largest
practical wavelength ranges may be more effective than the
conventional fitting technique employed here.

There are other, physical effects that might ultimately limit our
ability to measure $P(k)$ on very large scales.
One is the inhomogeneity of the UV background intensity
(and hence the photoionization rate $\Gamma$) due to the finite
number of sources (\cite{zuo92}).  Fluctuations in $\Gamma$
cause fluctuations in the \lya\ optical depth that are
unconnected to fluctuations in density.  
The short range of the \lya\ forest ``proximity effect''
relative to the mean separation of QSOs (see, e.g., \cite{bajtlik88};
\cite{bechtold94}) implies that fluctuations in $\Gamma$ are
much smaller than unity over most of space, but on large scales
the power induced by $\Gamma$ fluctuations might become 
comparable to the power in the density field itself.
Other subtle effects could become important for 
$\Delta z \ga 0.2$ ($2\pi/k \ga 60,000[1+z]^{-1}\;\kms$),
such as evolution of the UV background and evolution of the
density field itself.

Our method explicitly incorporates the assumption that primordial
fluctuations are Gaussian, as predicted in most inflationary models.
Figure~\ref{pkscdm} suggests that the recovered {\it shape} of
$P(k)$ is not particularly sensitive to this assumption; Gaussianization
is a useful, theoretically motivated tool for ``regularizing'' 
the observed absorption and thus reducing statistical fluctuations in
the $P(k)$ estimates, but because non-linear local transformations
do not alter the shape of the power spectrum at large scales,
we obtain a similar average $P(k)$ from our simulated spectra whether
or not we Gaussianize the flux.
We rely more heavily on the Gaussian assumption when we normalize $P(k)$,
since we employ PM simulations with random phase initial conditions.
Because we use the flux power spectrum, a variance measure, as our
criterion for normalizing $P(k)$, we might obtain similar results even
for other assumptions, but the sensitivity of the normalization to 
the Gaussian assumption can only be assessed quantitatively in the
context of a more explicit model for non-Gaussian primordial fluctuations.

The recovery of the mass power spectrum relies more generally
on the theoretical scenario for the \lya\ forest provided by
cosmological simulations and outlined in \S 2.1.
\lya\ forest observations also provide the means to
test this scenario, and to test the assumption of Gaussian fluctuations,
especially once the range of cosmological models to be considered
is narrowed by the $P(k)$ determination.
Traditional measures based on line-fitting,
such as the column density and $b$-parameter distributions,
provide one class of important tests.
Statistical methods that treat the spectrum as a continuous field rather
than a superposition of lines may ultimately prove more powerful,
since they are tied more directly to the quantities observed and
are better attuned to the physics of a continuous, fluctuating IGM
(see, e.g., \cite{miralda96}; \cite{cwhk97}; \cite{rauch97}; 
Weinberg et al., in preparation).
Observations of absorption along neighboring lines of sight
probed by QSO pairs and groups can also provide crucial 
tests, by constraining the sizes and geometry of the
absorbing structures (\cite{bechtold94a}; \cite{dinshaw94}; 
\cite{dinshaw95}; \cite{crotts97}).

Within the broad context of 
inflationary models for structure formation,
``free'' parameters include $\Omega_0$, $\Lambda_0$, $h$, $\Omega_b$,
the neutrino density parameter $\Omega_\nu$
(with the implied cold dark matter density parameter 
$\Omega_c=\Omega_0-\Omega_\nu-\Omega_b$), the primeval
spectral index $n$, the ratio of tensor-to-scalar contributions
to large angle CMB anisotropies, and the energy density of the relativistic
particle background (from CMB photons and light neutrinos, for example).
Particular regions within this parameter space are often
identified as named models, such as mixed dark matter, $\Lambda$CDM,
open CDM, or tilted CDM.  These models resolve the observational
conflicts that beset the most ``natural'' inflation model
($\Omega_0=1$, $n=1$, etc.) by appealing to different variations of
the fundamental physics --- e.g., by adjusting the matter content,
the vacuum energy, the space geometry, or the inflaton potential.
In combination with existing observational constraints from CMB
anisotropies and large scale structure, precise measurements of
the shape and amplitude of $P(k)$ at $z \sim 2-4$ can rapidly
shrink the viable regions of parameter space and thus rule out
or severely restrict many of these conceptually distinct models.
These measurements even have the potential to rule out the general
inflationary, adiabatic fluctuation scenario for the origin of
structure by revealing a radically different $P(k)$, though our 
results from Q1422+231 suggest that they will not.

The $P(k)$ constraints from the \lya\ forest complement the
observational constraints from the CMB and large scale structure
because they probe epochs between recombination and the present day
and because they respond differently to parameter variations.
Consider, for example, the CCDM model, which appears at least
marginally incompatible with the $P(k)$ inferred from Q1422+231
(Figure~\ref{pkq1422}).  It is well known that the CCDM model
is incompatible with the observed virial masses of rich galaxy
clusters, because it combines $\Omega_0=1$ with a high value of $\sigma_8$
(see, e.g., White \etal 1993).  However, our $P(k)$ measurement
challenges this model on different grounds, and the challenge would
also apply to a low-$\Omega_0$ model that has the same $P(k)$ amplitude
at $z=3$, even though such a model might pass the cluster test.
A difference from large scale structure constraints also arises
because the directly observable ``length'' units for \lya\ forest
studies are $\kms$, and the relation of these scales to $\hmpc$
at $z=0$ depends on the time evolution of the Hubble expansion
rate, and hence on $\Omega_0$ and $\Lambda_0$.  The constraints
from the \lya\ forest $P(k)$ will be especially powerful if they
can be extended to redshifts low enough that the fluctuation growth
rate and the redshift dependence of $H(z)$ start to differ measurably
from one cosmological model to another.  Since our tests indicate
that high spectral resolution and high signal-to-noise ratio are not
essential, the data from the HST Absorption Line Key Project
(\cite{bahcall93}; \cite{jannuzi97}) and other HST studies of 
the low-$z$ \lya\ forest may well be adequate for this purpose.  
The critical questions
are whether the relation between \lya\ optical depth and mass density
remains sufficiently tight as the universe evolves and whether large scale
fluctuations remain measurable with the much lower mean absorption
observed at low redshift (reflecting the lower value of $A$ in 
equation~[\ref{eqn:tau}]).
High resolution hydrodynamic simulations evolved to $z=0$ 
should soon provide the means to answer these questions.

The prospects for determining $P(k)$ at $z=2-4$ with existing QSO
absorption data are excellent.  (At $z=4$,
continuum determination may be a significant obstacle
because of the high mean opacity.)  There are already numerous
Keck HIRES spectra comparable in quality to the 
Q1422+231 spectrum that we have analyzed here.
Furthermore, our tests show that spectra of moderate resolution
and signal-to-noise ratio are adequate for $P(k)$ determinations, and
since one wants as many spectra as possible in order to beat down
cosmic variance and achieve high statistical precision over a range of
redshifts, the large existing samples of ``pre-Keck'' spectra
(e.g., \cite{bechtold94}) may be even better suited to this purpose.
For future programs designed specifically for power spectrum studies,
observations targeting groups of QSOs with transverse separations
of several to several 10's of comoving $\hmpc$ (e.g., \cite{williger96})
might be especially valuable, as cross-correlating spectra along
parallel lines of sight rapidly multiplies the number of baselines
available for estimating $P(k)$.  With a sufficiently large sample,
one could even use the angular dependence of $P(k)$ to 
constrain spacetime geometry (\cite{alcock79}; \cite{matsubara96}; 
\cite{ballinger97}; \cite{popowski97}).
The Sloan Digital Sky Survey, which will obtain $\sim 10^5$ QSO spectra
with $2.5\AA$ resolution and typical S/N$\sim 10-20$,
may eventually provide the ideal data set for such a study.

The promise of the method described in this paper arises from the happy
coincidence between the excellent observational data on the \lya\ forest
and the simple physical interpretation of these data that has emerged
from cosmological simulations.  Because the \lya\ forest arises
primarily in the diffuse, smoothly fluctuating IGM, this route to $P(k)$
sidesteps the uncertainties in theoretical modeling of galactic star formation,
feedback, galaxy mergers, and so forth, which inevitably affect 
interpretations of large scale structure data.  Determination of the
mass power spectrum in the high redshift universe now looks to be within
relatively easy reach.

\acknowledgments
We thank A. Songaila and L. Cowie for providing the Q1422+231 spectrum
and J. Miralda-Escud\'e for helpful discussions and for the program
used to generate the Voigt-profile line model used in Figure~\ref{pklinen-1}.
This work was supported by NASA Astrophysical Theory Grants
NAG5-2864, NAG5-3111, NAGW-2422, NAG5-2793, and NAG5-3922,
by NASA Long-Term Space Astrophysics Grant NAG5-3525,
and by the NSF under grants ASC93-18185 and the Presidential
Faculty Fellows Program.
The SPH simulations were performed at the San Diego Supercomputer Center.

\vfill\eject

\end{document}